\documentclass[12pt]{article}
\pdfoutput=1
\usepackage{epsf,amsfonts,amssymb,epsfig,amsmath,mathtools,graphics,slashed}
\usepackage{hyperref,hep,graphicx,subfig}
\usepackage{rotating}

\newcommand{\nn}{\notag \\}

\newcommand{\dimdef}{{\Lambda}}

\makeatletter

\@addtoreset{equation}{section}
\makeatother

\newcommand{\dd}{\mathrm{d}}

\newcommand{\bb}{\bar{\beta}}

\begin{document}

\begin{titlepage}

\vfill

\begin{flushright}
Imperial/TP/2018/JG/02\\
\end{flushright}

\vfill

\begin{center}
   \baselineskip=16pt
   {\Large\bf Susy Q and spatially modulated\\ deformations of ABJM theory}
  \vskip 1.5cm
Jerome P. Gauntlett and Christopher Rosen\\
     \vskip .6cm
             \begin{small}\vskip .6cm
      \textit{Blackett Laboratory, 
  Imperial College\\ London, SW7 2AZ, U.K.}
        \end{small}\\
                       \end{center}
\vfill

\begin{center}
\textbf{Abstract}
\end{center}
\begin{quote}
Within a holographic framework we construct supersymmetric Q-lattice (`Susy Q') solutions that
describe RG flows driven by supersymmetric and spatially modulated deformations of the dual CFTs. We focus on a specific $D=4$ supergravity model which arises as a consistent KK truncation of $D=11$ supergravity
on the seven sphere that preserves $SO(4)\times SO(4)$ symmetry. The
Susy Q solutions are dual to boomerang RG flows, flowing from ABJM theory in the UV, deformed
by spatially modulated mass terms depending on one of the spatial directions, back to the ABJM vacuum in the far IR. For large enough deformations the boomerang flows approach the well known Poincar\'e invariant RG dielectric flow. The spatially averaged energy density vanishes for the Susy Q solutions.

\end{quote}

\vfill

\end{titlepage}

\section{Introduction}

Holography provides a powerful framework for analysing what happens to strongly coupled
field theories that have been deformed by spatially modulated operators. Indeed, by solving the relevant bulk gravitational
equations of motion one can deduce the entire RG flow from the UV to the IR that is induced by the deformation.
An interesting class of such RG flows are the boomerang RG flows \cite{Chesler:2013qla,Donos:2014gya,Donos:2016zpf,Donos:2017ljs,Donos:2017sba}, which start from a fixed point in the UV and end up at exactly the same fixed point in the IR. 

Several of these boomerang flows have been studied within the framework of
Q-lattice constructions \cite{Donos:2013eha}, which leads to important technical simplifications in solving the bulk equations.
Such constructions require the bulk gravitational theory to admit a global symmetry and this is used
to provide an ansatz for the bulk fields whereby, effectively, the spatial dependence
of the fields is solved exactly. The full equations of motion then boil down to solving a system
of ordinary differential equations that just depend on the holographic radial coordinate.

Boomerang RG flows for CFTs in $d$ spacetime dimensions that are driven by spatially modulated operators that depend on
a single Fourier mode, $k$, are parametrised by a single dimensionless parameter 
$\dimdef/k^{d-\Delta}$ where $\Delta$ is the scaling dimension of the
operator, $\mathcal{O}$, in the CFT. For small values of this parameter, when $\mathcal{O}$ is a relevant operator\footnote{For marginal operators
$\mathcal{O}$, when $\Delta=d$, the RG flows can be boomerang flows or flows to other behaviour in the IR; see \cite{Donos:2016zpf} for explicit examples.} with
$\Delta<d$
one can show that the RG flow is a boomerang flow, flowing from an $AdS$ vacuum in the UV to the same $AdS$ vacuum in the IR.
by perturbatively solving the bulk equations. For large values of $\dimdef/k^{d-\Delta}$, however,
one needs to numerically solve the system of ODEs. In some examples, the boomerang flow persists but in others
there can be a quantum phase transition at some critical value of $\dimdef/k^{d-\Delta}$ leading to new IR behaviour.

If the boomerang RG flow does exist for arbitrary large values of $\Lambda/k^{d-\Delta}$ then 
as $\Lambda/k^{d-\Delta}\to\infty$ the RG flow approaches the Poincar\'e invariant RG flow that is driven by the relevant operator, $\mathcal{O}$, with $k=0$, before returning to the $AdS$ vacuum in the far IR. Thus, these boomerang flows exhibit an interesting intermediate regime, dominated by the IR behaviour
of the Poincar\'e invariant RG flow, that generically\footnote{It is not automatic that this is the case; see \cite{Donos:2017sba} for more details.} imprints itself on various
observables, such as spectral density functions and entanglement entropy \cite{Donos:2017sba}. In some models
the Poincar\'e invariant RG flow has a singular behaviour in the IR and in these cases, the boomerang RG flows
can be viewed as a novel mechanism to resolve this singularity. 

The starting point for the results presented here was the observation that if the Poincar\'e invariant RG flow is supersymmetric then the associated boomerang flows, if they exist for large enough deformations, should exhibit approximate intermediate supersymmetry. In studying this in more detail, we found an interesting Q-lattice construction in which the entire boomerang RG flows, for arbitrary values of the deformation parameter
$\dimdef/k^{d-\Delta}$, preserve some supersymmetry.
Naturally enough, we christen Q-lattices that preserve supersymmetry `Susy Q'.

We begin by examining general Susy Q constructions in the context of $\mathcal{N}=1$ supergravity in $D=4$
with a single chiral field. If the model has a constant superpotential then
one can construct a Susy Q ansatz that is anisotropically spatially modulated in just one of the two spatial
directions of the dual field theory and preserves 1/4 of the supersymmetry. We then focus on
a specific top-down model that arises as a consistent KK truncation of $D=11$ supergravity on $S^7$
that preserves $SO(4)\times SO(4)$ symmetry \cite{Cvetic:1999au}. After uplifting the Susy Q solutions to $D=11$
we obtain boomerang RG flows driven by specific spatially modulated mass terms in the dual ABJM theory 
\cite{Aharony:2008ug}
associated with operators of dimension $\Delta=1$ and $2$. For large deformations the
boomerang flows have an intermediate regime which approaches the Poincar\'e invariant dielectric flows
studied in \cite{Pope:2003jp}.
Interestingly, and independently of this
work, it was recently shown that spatially modulated mass terms for ABJM, depending on one of the
spatial coordinates can preserve supersymmetry in \cite{Kim:2018qle}, and our work provides the gravity dual
for a specific example.
Another gravity dual is provided by the supersymmetric Janus solutions presented in \cite{DHoker:2009lky,Bobev:2013yra}.

We also calculate the stress tensor for the Susy Q solutions. One 
interesting feature is that the stress tensor is spatially modulated. 
In a generic Q-lattice construction, the metric is spatially homogeneous and hence, with
standard holographic boundary terms, this implies a spatially homogeneous stress tensor. 
In our set-up, however, supersymmetry demands that we have boundary terms such that the real and imaginary parts of
the bulk complex scalar field are dual to operators with dimension $\Delta=1$ and $2$, respectively.
In particular, this requires that we add boundary terms which break the bulk global symmetry being
used in the Q-lattice construction and this leads to the stress tensor having non-trivial dependence on
the spatial coordinates. Another interesting feature is that the stress tensor for the Susy Q solutions
has zero average energy density, $\langle \overline{\mathcal{T}^{tt}}\rangle=0$, where the bar refers to taking the average over a spatial period. We will see that this is associated with a novel way to preserve supersymmetry, utilising the spatial periodicity of the configuration.

Top down, isotropic boomerang flow solutions were found in \cite{Donos:2017ljs}
using a Q-lattice ansatz. In appendix \ref{sec:HRGs} we calculate the spatially modulated 
stress tensor for these solutions and find
$\langle \overline{\mathcal{T}^{tt}}\rangle\ne 0$, as expected, since the solutions do not preserve supersymmetry.
An interesting feature of these isotropic 
boomerang flows is that for large enough deformations, when the flows approach
the Poincar\'e invariant RG flow, they also approach a second intermediate scaling regime with
hyperscaling violation. This latter regime is determined not by a solution to the equations of motion themselves
but to those of an auxiliary gravitational theory whose equations of motion approximately agree
when the scalar field becomes large. 
While this phenomenon is rather natural from a gravitational point of view, it is less so
from the field theory point of view and deserves further study. 
It is therefore natural to investigate if a similar thing happens for the Susy Q solutions constructed here. In appendix \ref{appdee} we show that the obvious auxiliary gravitational theory has a novel
$AdS_3\times\mathbb{R}$ solution with the remarkable property that the value of the scalar field is not fixed.
However, it turns out that this $AdS_3\times\mathbb{R}$ geometry does not play a role in the Susy Q boomerang RG flows which we construct. Appendix \ref{appdee} also discusses a general class of gravity theories that
admit similar and novel $AdS_{D-n}\times\mathbb{R}^n$ solutions, breaking translations in the
$\mathbb{R}^n$ directions,
which would be interesting to explore further.

\section{Susy Q}
In this section we describe a general construction of supersymmetric, anisotropic Q-latices. We work within the framework of $\mathcal{N}=1$ supergravity in $D=4$ spacetime dimensions coupled to a single chiral multiplet
(see appendix \ref{sec:Nis1con} for more details).
The bosonic part
of the action is given by
\begin{equation}\label{eq:Sg0}
S= \int \dd^4 x \sqrt{-g}\Big(R - G\partial_\mu z \partial^\mu \bar{z}-\mathcal{V}\Big)\,.
\end{equation}
The complex scalar $z$ parametrises a K\"ahler manifold with $G=2\partial\bar \partial\mathcal{K}$
where $\mathcal{K}(z,\bar z)$ is the K\"ahler potential. The potential $\mathcal{V}(z,\bar z)$ is given by
\begin{equation}
\mathcal{V} = 4 G^{-1}\partial\mathcal{W}\bar \partial\mathcal{W}-\frac{3}{2}\mathcal{W}^2\,,\qquad
\mathcal{W} =-e^{\mathcal{K}/2}|W|\,,
\end{equation}
where $W$ is the superpotential which is a holomorphic function of $z$.

For a Q-lattice construction \cite{Donos:2013eha}
we require that the model admits a global abelian symmetry. Assuming that this 
acts as a constant phase rotation of the field $z$, we demand that $G$ and $\mathcal{V}$, which are functions
of both $z$ and $\bar z$, in general, are functions of $|z|$ only.
We can then consider the anisotropic Q-lattice ansatz, consistent with equations of motion, given by\footnote{Note that we can also replace $z$ with $z=\rho e^{ikx+i\theta}$ for some constant $\theta$ without changing any of the formulae below. When $k\ne 0$ we can absorb $\theta$ into
a shift of the $x$ coordinate. However, when $k=0$ the value of $\theta$ can play an important role in the context of consistent KK truncations when uplifting
to obtain solutions in higher dimensions.} 
\begin{align}\label{Susyqansatz}
ds^2&=e^{2A}(-dt^2+dy^2)+e^{2V} dx^2 +N^2dr^2\,,\nn
z&=\rho e^{ikx}\,,
\end{align}
with $A$, $V$, $N$ and $\rho$ all functions of the radial coordinate $r$ only. Without loss of generality
we will assume $k>0$ in the sequel.
Translations in the $x$ direction
are broken by this ansatz, as is the global $U(1)$ symmetry, but a diagonal combination of the two symmetries is preserved. 

We proceed by taking a gauge for the K\"ahler potential with 
$\mathcal{K}=\mathcal{K}(|z|)$. Then the only non-vanishing component of the K\"ahler
connection one-form, defined by $\mathcal{A}_\mu=\frac{i}{6}(\partial\mathcal{K}\partial_\mu z-\bar\partial\mathcal{K}\partial_\mu \bar z)$, is the $x$ component with 
$\mathcal{A}_x=-1/6\rho k\mathcal{K}'$. In order to find supersymmetric solutions,
satisfying projections on the Killing spinors given below, we now restrict to
models with constant superpotential, which, we can take to be real
\begin{align}
W=constant\in \mathbb{R}\,.
\end{align}
Using the analysis of appendix \ref{sec:Nis1con} we then obtain the following set of BPS equations 
\begin{align}
\label{bpsgen}
N^{-1}\rho'-ke^{-V}\rho+\frac{1}{2}e^{\mathcal{K}/2}G^{-1}\mathcal{K}' W&=0\,,\nn
{N}^{-1}A'-\frac{1}{2}e^{\mathcal{K}/2}W&=0\,,\nn
N^{-1}V'+\frac{1}{2}e^{-V}k\rho\mathcal{K}'-\frac{1}{2}e^{\mathcal{K}/2}W&=0\,.
\end{align}
It is straightforward to show that any solution to these BPS equations automatically
solves the full equations of motion. 

Our primary interest is within the context of holography and so we assume that the model admits
a vacuum $AdS_4$ solution. We also assume that the two real scalar fields in $z$ are dual to 
relevant (or possibly marginal) operators in the dual conformal field theory. Setting $k=0$, the above ansatz can be used to construct 
supersymmetric solutions that describe a Poincar\'e invariant RG flow from the deformed CFT in the UV to some other behaviour in the IR. The latter could be, for example, another $AdS_4$ fixed point, but there
are many other possibilities too. When $k\ne 0$ the supersymmetric solutions describe the RG flows associated
to deformations of the CFT in the UV that also break translations in the $x$ direction.

When $k=0$ the Killing spinors, $\hat\epsilon$, are given by
\begin{align}\label{ksrad}
\hat\epsilon=e^{A/2}\eta,\qquad \Gamma^{\hat r}\eta=-\eta\,,
\end{align}
where $\eta$ is a constant Majorana spinor.
The projection implies that the superconformal symmetries of the dual field theory are broken but the Poincar\'e
supersymmetries are preserved, as expected for the RG flow. When $k\ne0$ we also need to impose 
an additional projection on the Killing spinor given by
\begin{align}\label{ksbeg}
\Gamma^{\hat t \hat y}\eta=-\eta\,,
\end{align}
which breaks 1/2 of the Poincar\'e supersymmetries. It is also worth pointing out that the Killing vector 
that can be constructed from the Killing spinor
bi-linear via $(\bar{\hat\epsilon}\Gamma^\mu \hat \epsilon)\partial_\mu$
is null and proportional to  $\partial_t+\partial_y$.

\section{Susy Q in ABJM theory}\label{sfthy}
We now focus on a specific $D=4$ supergravity theory with action given by
\begin{equation}\label{eq:Sg2bulk}
S= \int\dd^4 x \sqrt{-g}\Big(R - \frac{2}{(1-|z|^2)^2}\partial_\mu z \partial^\mu \bar{z}
+\frac{2(3-|z|^2)}{1-|z|^2}\Big)\,.
\end{equation}
In particular, we have $e^{\mathcal{-K}/2}=2^{3/2}(1-|z|^2)^{1/2}$ and $W=2^{5/2}$.
This arises as a consistent KK truncation of $D=11$ supergravity on $S^7$ and hence any solution 
can be uplifted on an $S^7$ to obtain a solution of $D=11$ supergravity, and hence is of relevance to 
the dual ABJM field theory \cite{Aharony:2008ug}. 
Starting with the maximal $\mathcal{N}=8$ $SO(8)$ gauged supergravity \cite{deWit:1981sst}
we can further truncate
to $\mathcal{N}=4$ $SO(4)$ gauged supergravity \cite{Das:1977pu}, 
whose bosonic sector is given by \eqref{eq:Sg2bulk}, after setting the gauge fields to zero. The model can also be obtained as a truncation of the $\mathcal{N}=2$ STU gauged supergravity
theory \cite{Cvetic:1999xp,Cvetic:2000tb}, as we discuss further in appendix \ref{conkk}. The formulae for the uplifted $D=11$ Susy Q solutions,
which preserve $SO(4)\times SO(4)$ symmetry, can be found using the results of \cite{Cvetic:1999au}, and are presented in section \ref{seceleven}. We write the real and imaginary parts of $z$ as
\begin{align}
z=\mathcal{X}+i\mathcal{Y}\,,
\end{align}
and, without loss of generality, take $\mathcal{X}$ to be one of the 35 scalars of $\mathcal{N}=8$ gauged supergravity
and $\mathcal{Y}$ to be one of the 35 pseudoscalars.

The $AdS_4$ vacuum solution\footnote{Note that for convenience we have set the radius of the $AdS_4$ space to unity and we have also set $16\pi G=1$.} with $z=0$, uplifts to the $D=11$ $AdS_4\times S^7$ vacuum solution.
In this vacuum $\mathcal{X}$ and $\mathcal{Y}$ have mass squared equal to minus two and hence
we can impose boundary conditions so that they are dual to operators, $\mathcal{O}_\mathcal{X}$ and
$\mathcal{O}_\mathcal{Y}$, with scaling dimensions $\Delta=1$ or $\Delta =2$. However, supersymmetry demands \cite{Breitenlohner:1982jf}
that we must choose boundary conditions, which we discuss in appendix \ref{holren}, so that 
\begin{align}
\Delta_\mathcal{X}=1\,,\qquad \Delta_\mathcal{Y}=2\,.
\end{align}

For this particular model the BPS equations \eqref{bpsgen} are given by
\begin{align}\label{eq:bps1}
0 = & \, r A' -\frac{1}{\sqrt{1-\rho^2}}\,,\nn
0 = &\, rV'-\frac{1}{\sqrt{1-\rho^2}}+ke^{-V}\frac{\rho^2}{1-\rho^2}\,,\nn
0 = & \,  r\frac{\rho'}{\rho} -ke^{-V}+\sqrt{1-\rho^2}\,,
\end{align}
where we have now chosen, for convenience,
\begin{align}
N=\frac{1}{r}\,.
\end{align}
We showed in the previous section that these solutions preserve one supersymmetry
in $\mathcal{N}=1$ supergravity for $k\ne 0$. 
We show in appendix \ref{appb} that such solutions preserve eight supersymmetries in the context
of $\mathcal{N}=8$ supergravity.

\newcommand{\dfone}{\Lambda}
\newcommand{\dftwo}{l_{(2)}}

We are interested in solutions to the BPS equations that asymptotically approach $AdS_4$ in the UV, which we take to be located at $r\to\infty$.
Using the second order equations of motion
we can construct the following expansion as $r\to\infty$, 
\begin{align}\label{eq:bpsFOs1}
e^{2A} & = r^2 -\frac{1}{2}{\dfone}^2 + M\frac{1}{r} + \ldots\,,\nn
e^{2V}& =\, r^2 -\frac{1}{2}{\dfone}^2- (2M+\frac{8}{3}{\dfone}{\dftwo})\frac{1}{r} + \ldots\,,\nn
\rho& =\, {\dfone}\frac{1}{r} + {\dftwo}\frac{1}{r^2}+\ldots\,,
\end{align}
with the higher order terms determined by $M$, ${\dfone}$, and ${\dftwo}$.
For solutions of the BPS equations \eqref{eq:bps1} we have the additional constraints
\begin{equation}\label{eq:bpsFOs}
{\dftwo} = -k {\dfone},\qquad M = \frac{2}{3}k {\dfone}^2\,.
\end{equation}

Using some results on holographic renormalisation, summarised in appendix \ref{holren}, we deduce that the field theory has non-trivial
sources for the scalar operators with
\begin{equation}\label{eq:defstext}
\mathcal{X}_s = -4{\dftwo}\cos kx = 4k {\dfone}\cos k x\,,
\qquad 
{\mathcal{Y}}_s = {\dfone}\sin kx\,, 
\end{equation}
and, furthermore, we also deduce the following expectation values
\begin{align}\label{vevs}
\langle\mathcal{T}^{tt} \rangle = \,-\langle\mathcal{T}^{yy} \rangle=&\,-3M-4{\dfone} {\dftwo}\sin^2 kx = -2k\,{\dfone}^2\cos 2k x\,,\nn
\langle\mathcal{T}^{xx} \rangle = &\,  -2(3M+{\dfone}{\dftwo}(3+\cos 2kx)) = 4k\, {\dfone}^2\cos^2 k x\,,\nn
\langle  \mathcal{O}_{\mathcal{X}}\rangle = &\, {\dfone}\cos k x\,,\nn
\langle   \mathcal{O}_{\mathcal{Y}}\rangle = &\,  4{\dftwo}\sin kx = -4k\,{\dfone}\sin k x\,.
\end{align}
In \eqref{eq:defstext}, \eqref{vevs} the first expressions are valid for general solutions to the equations of motion and the final expressions valid for solutions to the BPS equations.
One can check that these  satisfy the Ward identities
\begin{align}\label{eq:Ward1text}
 \partial^i\langle \mathcal{T}_{ij}\rangle  = &\, \langle  \mathcal{O}_{\mathcal{X}}\rangle\partial_j\mathcal{X}_s+ \langle  \mathcal{O}_{\mathcal{Y}}\rangle\partial_j\mathcal{Y}_s\,,\nn
 \langle \mathcal{T}^i\,_i\rangle = & \, (3-\Delta_\mathcal{X})\, \langle \mathcal{O}_{\mathcal{X}}\rangle\mathcal{X}_s+(3-\Delta_\mathcal{Y})\, \langle  \mathcal{O}_{\mathcal{Y}}\rangle \mathcal{Y}_s\,.
\end{align}
Also notice that for solutions to the BPS equations,
the dimensionless quantities $\langle \mathcal{T}^{ij}\rangle/k^3$, $\langle  \mathcal{O}_{\mathcal{X}}\rangle/k$
and $ \langle  \mathcal{O}_{\mathcal{Y}}\rangle/k^2$ 
all depend on the deformation parameter $\Lambda$ via the dimensionless combination $\Lambda/k$. 

\begin{figure}
\centering
\includegraphics[scale=0.33]{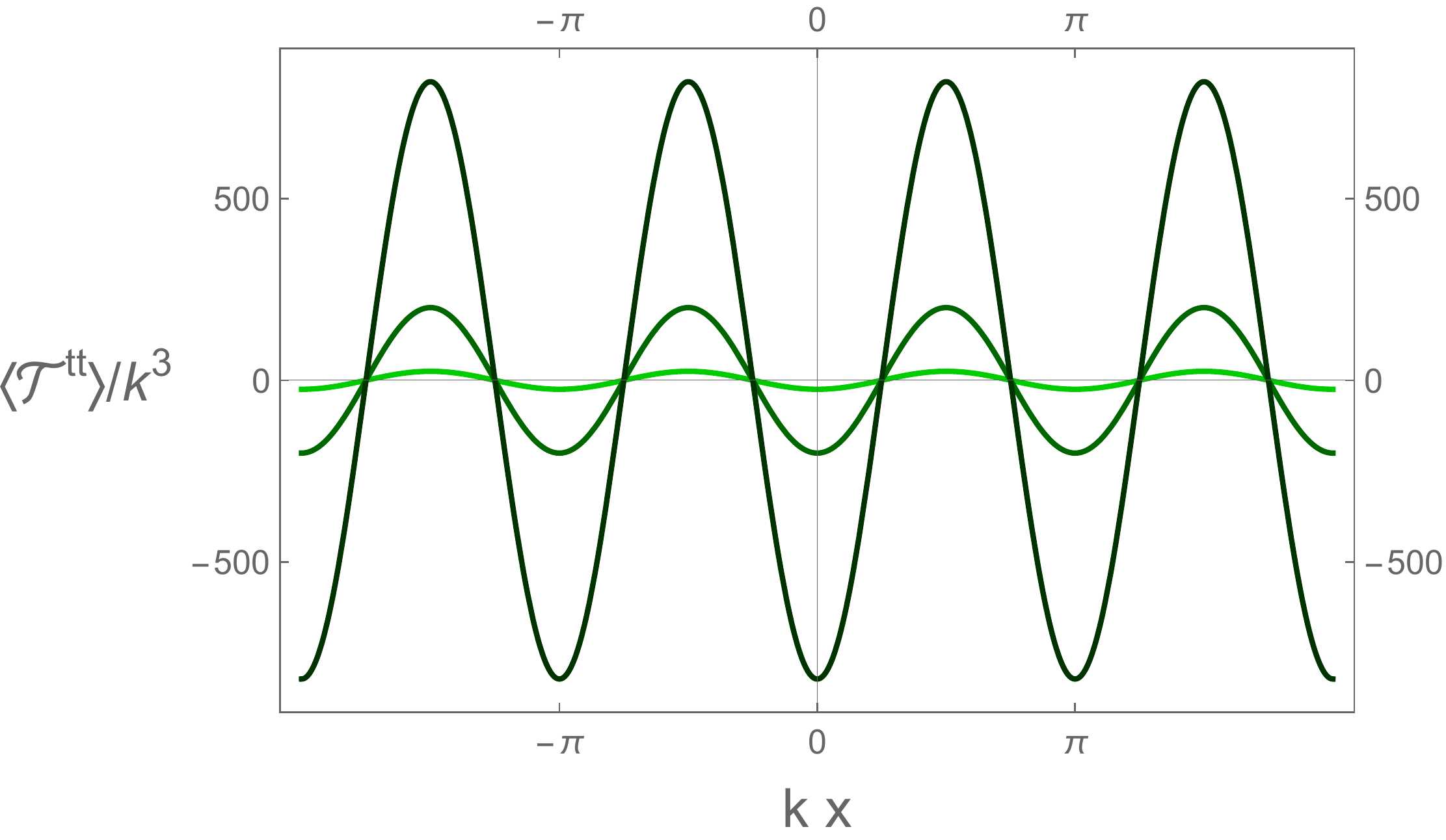}
\includegraphics[scale=0.33]{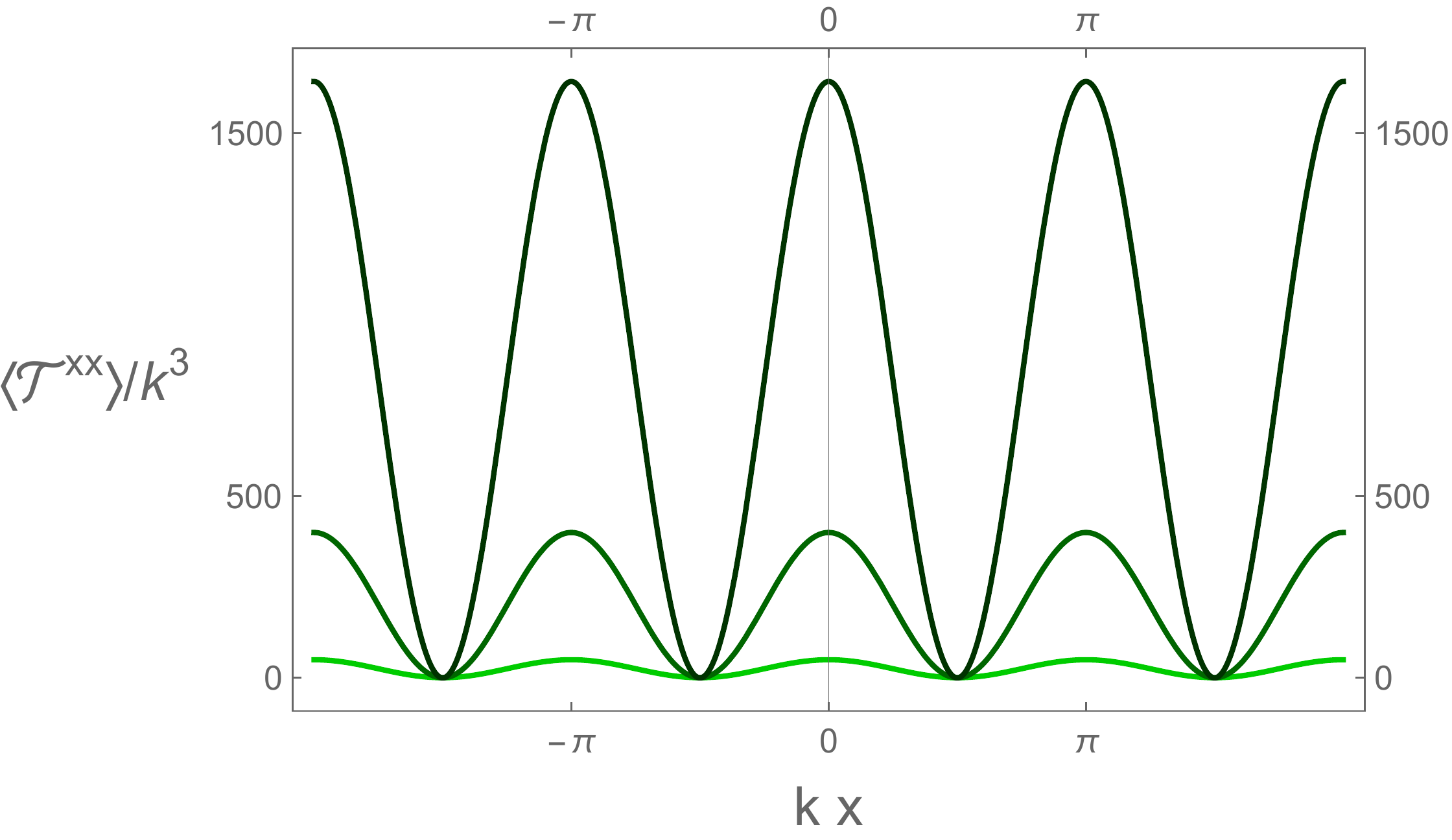}
\caption{\label{fig:Taniso} The one-point functions of the stress energy tensor $\langle\mathcal{T}^{tt}\rangle = - \langle\mathcal{T}^{yy}\rangle$ and $\langle\mathcal{T}^{xx} \rangle$ for boomerang RG flows driven by deformations with $\Lambda/k\approx 3.5,\, 10.0,\,20.3$ (lightest to darkest).  The spatially averaged energy density vanishes in these flows, $\langle\overline{\mathcal{T}^{tt}}\rangle=0$.}
\end{figure}

We now discuss some interesting features concerning the stress tensor which is plotted in
figure \ref{fig:Taniso} for various deformation parameters $\Lambda/k$.
First, 
it has non-trivial dependence on the spatial coordinates. Recall that for generic
Q-lattice constructions, with boundary terms that preserve the global symmetry associated with the Q-lattice construction, the stress tensor is spatially homogeneous. In particular, the bulk metric is spatially homogeneous in the field theory directions and furthermore the extrinsic curvature is as well. In order to obtain
a spatially inhomogeneous stress tensor it is necessary that the boundary terms break
the global symmetry. Indeed
the spatial dependence that
we find here is precisely because we have imposed alternate quantisation boundary conditions for the scalar field
$\mathcal{X}$ in order that $\Delta_\mathcal{X}=1$.

Second, if we average over a spatial period in the $x$ direction, we have
$\langle  \overline{\mathcal{ T}^{tt}} \rangle=\langle   \overline{\mathcal{ T}^{yy}} \rangle=0$
and $\langle   \overline{\mathcal{T}^{xx}}\rangle=2k{\dfone}^2$. The fact that the average
energy density is zero follows from supersymmetry. Indeed from the first expression for
$\langle\mathcal{T}^{tt} \rangle$ in \eqref{vevs} one sees that the average energy density is generically non-zero for Q-lattice solutions of the second order equations of motion. Furthermore, it is also worth emphasising
that $\langle  \overline{\mathcal{T}^{tt}} \rangle$ =0 requires not only the BPS equations but also
the boundary conditions associated with the alternate quantisation of $\mathcal{X}$, both of which are required
for supersymmetry.

\subsection{The Poincar\'e invariant dielectric flow}\label{pinvflow}
When $k=0$, there is a solution to the BPS equations \eqref{eq:bps1} 
that gives a Poincar\'e invariant RG flow first discussed in \cite{Pope:2003jp}. This flow preserves sixteen supersymmetries when embedded in $\mathcal{N}=8$ gauged supergravity, and can be written in our conventions as
\begin{equation}\label{poincflowexp}
e^{2A}=e^{2V}=  r^2\left(1-\frac{\mu^2}{r^2} \right)^2, \qquad z = \frac{2\mu r}{r^2+\mu^2}e^{i\theta}\,,
\end{equation}
for any constant $\theta$. The value of $\theta$ plays an interesting role when the solutions are uplifted to
$D=11$ supergravity\footnote{These $D=11$ solutions are generalised in \cite{Bena:2004jw,Lin:2004nb}.}. Indeed, as $\theta$ varies from $0$ to $\pi/2$, 
the solution rotates between a holographic description of a purely Coulomb branch RG flow, associated with a distribution of membranes, in which the 
operator $\mathcal{O}_\mathcal{X}^{\Delta=1}$ has condensed, and a `dielectric' RG flow, with membranes puffing up into fivebranes, driven by a supersymmetric source for the $\mathcal{O}_\mathcal{Y}^{\Delta=2}$ operator. 

The Ricci scalar diverges for these solutions at $r = \mu$. When $\theta =0$, for example, this curvature singularity can be simply understood from the eleven dimensional perspective in terms of the distribution of membranes \cite{Pope:2003jp}, much like the $D=5$ analogue \cite{Freedman:1999gk}. 

From the perspective of applied holography, a particularly noteworthy feature of this solution
is that operators in the dual ABJM phase are generically gapped at low frequencies. To illustrate this
we can consider linearized fluctuations of a massless scalar in this background of the form
\begin{equation}
\delta h(t,r) = h(r)e^{-i\omega t}\,.
\end{equation}
These could be, for example, fluctuations of the transverse traceless modes of the metric. The linearised equation can be solved exactly for $h$ to get
\begin{equation}
h = \left(\frac{r-\mu}{r+\mu} \right)^{\sqrt{1-\frac{\omega^2}{4\mu^2}}}\left[\frac{r^2+\mu^2 + 2 r\mu\sqrt{1-\frac{\omega^2}{4\mu^2}}}{r^2-\mu^2}\right]\,.
\end{equation}
Here we have demanded that we have ingoing boundary conditions for $\omega>2\mu$ (which 
also means, as it turns out, choosing the most regular solutions as we approach the singularity at $r\to \mu$).
This gives rise to a retarded Green's function for the dual dimension three operator of the form
\begin{equation}\label{retgf}
G^R(\omega) \propto \omega^2\mu\sqrt{1-\frac{\omega^2}{4\mu^2}}.
\end{equation}
Crucially, this correlation function is purely real for frequencies $\omega \le 2\mu$. Since the spectral function for this operator is proportional to the imaginary part of the two point function, the spectral weight is gapped for energies below $2\mu$. Other bosonic and fermionic probes from the dual field theory exhibit similar behaviour 
for these flows \cite{DeWolfe:2014ifa,Kiritsis:2015oxa}.

The presence of the gap is associated with the detailed way in which the background solution is becoming singular in the IR.
In the next section we will construct supersymmetric boomerang RG flow solutions which, for large deformations, approach
this Poincar\'e invariant flow before flowing back to the $AdS_4$ vacuum in the far IR. These boomerang solutions
both regulate the singularity of the solution as well as close the gap in the spectral functions.

\subsection{Holographic boomerang RG flows}

We now show that the BPS equations \eqref{eq:bps1} with $k\ne 0$ admit boomerang solutions which flow from
the $AdS_4$ ABJM vacuum in the UV and then return to the same vacuum in the far IR. We we will first exhibit such solutions, analytically, at leading order in a perturbative expansion with respect to the dimensionless deformation parameter ${\Lambda}/{k}$.
We will then show that boomerang flows also exist for large deformations by solving the BPS equations numerically,
and show that the solutions exhibit a region that approaches the Poincar\'e invariant solution discussed in
the last sub-section.

An interesting observable of the boomerang flows is the 
`index of refraction' which measures the renormalisation of relative length scales in the UV 
and IR as a ratio of the coordinate speeds of light there. For the anisotropic flows we are considering,
which preserve Poincar\'e invariance in the $t,y$ plane,
the only non-trivial index of refraction is in the $x$ direction, $n_x$. If we define $\chi=e^{A-V}$ then 
$n_x\equiv\frac{\chi_{UV}} {\chi_{IR}}$. From the BPS equations, we have
\begin{equation}
\frac{\dd \log \chi}{\dd \log r} = {k}e^{-V}\left(\frac{\rho^2}{1-\rho^2}\right) \ge 0\,,
\end{equation}
from which we deduce that $n^x \ge 1$. An analogous result was proven for some isotropic boomerang flows in \cite{Donos:2017sba} and it seems likely that this is a general result in holography.

\subsubsection{Perturbative anisotropic boomerang RG flows}

It is straightforward to show that for small values of $\Lambda/k$ the 
BPS equations \eqref{eq:bps1} admit boomerang RG flow solutions. 
Working to quadratic order in $\Lambda/k$, one can develop the expansion
\begin{align}
\label{eq:pBoomAn}
\rho &=\frac{\Lambda}{r}e^{-k/r}+\dots \,,\qquad e^{2A} = r^2\left[1+\frac{\Lambda ^2}{4k^2}\left(-1+\frac{ (2 k+r)}{  r}e^{-2 k/r}\right)+\dots \right]\,,\nn
e^{2V}&= r^2\left[1+\frac{\Lambda^2}{4k^2}\left(1-\frac{ \left(4 k^2+2 k r+r^2\right)}{ r^2}e^{-2 k/r}\right)+\dots \right]\,.
\end{align}
This solution asymptotes to the unit radius $AdS_4$ in the UV ($r\to\infty$), with the correct boundary conditions
given in \eqref{eq:bpsFOs1}, \eqref{eq:bpsFOs}. Moreover, it also asymptotes to the same unit radius $AdS_4$
in the IR ($r\to 0$).
We can immediately read off the index of refraction in the $x$ direction and we find
\begin{equation}\label{pertnx}
n^x = 1+\frac{1}{4}\left(\frac{\Lambda}{k}\right)^2+O\left(\frac{\Lambda}{k}\right)^3\,.
\end{equation}

\subsubsection{Non-perturbative anisotropic boomerang RG flows}
In order to study the RG flows for values of $\Lambda/k$ outside the perturbative regime, we numerically integrate the BPS equations. A simple and effective method is to use a perturbative boomerang solution (\ref{eq:pBoomAn}) to seed a shooting algorithm which integrates the background fields towards the UV.

\begin{figure}
\centering
\includegraphics[scale=0.38]{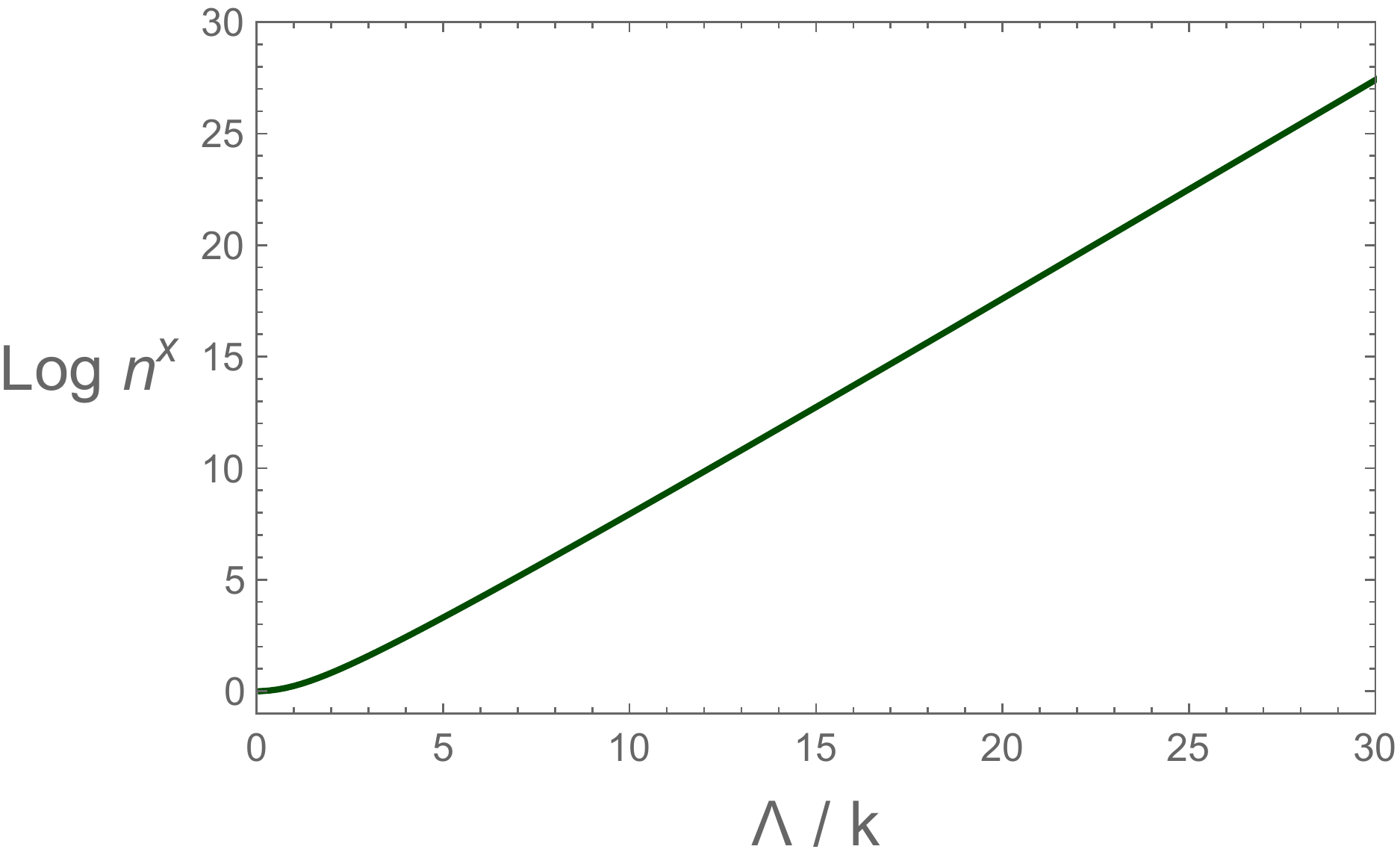}
\caption{\label{fig:boomSur} The index of refraction, $n_x$, for the BPS boomerang flows as a function of the dimensionless deformation parameter $\Lambda/k$.}
\end{figure}

For this model we find that the boomerang RG flows seem to exist for arbitrarily large values of $\Lambda/k$.
In figure \ref{fig:boomSur} we plot the value of the index of refraction as a function of $\Lambda/k$. 
The results agree with the perturbative result \eqref{pertnx} for small $\Lambda/k$. The 
figure also shows that for very large deformations, the index of refraction grows approximately exponentially with increasing $\Lambda/k$. 

\begin{figure}[!h]
\centering
\includegraphics[scale=0.35]{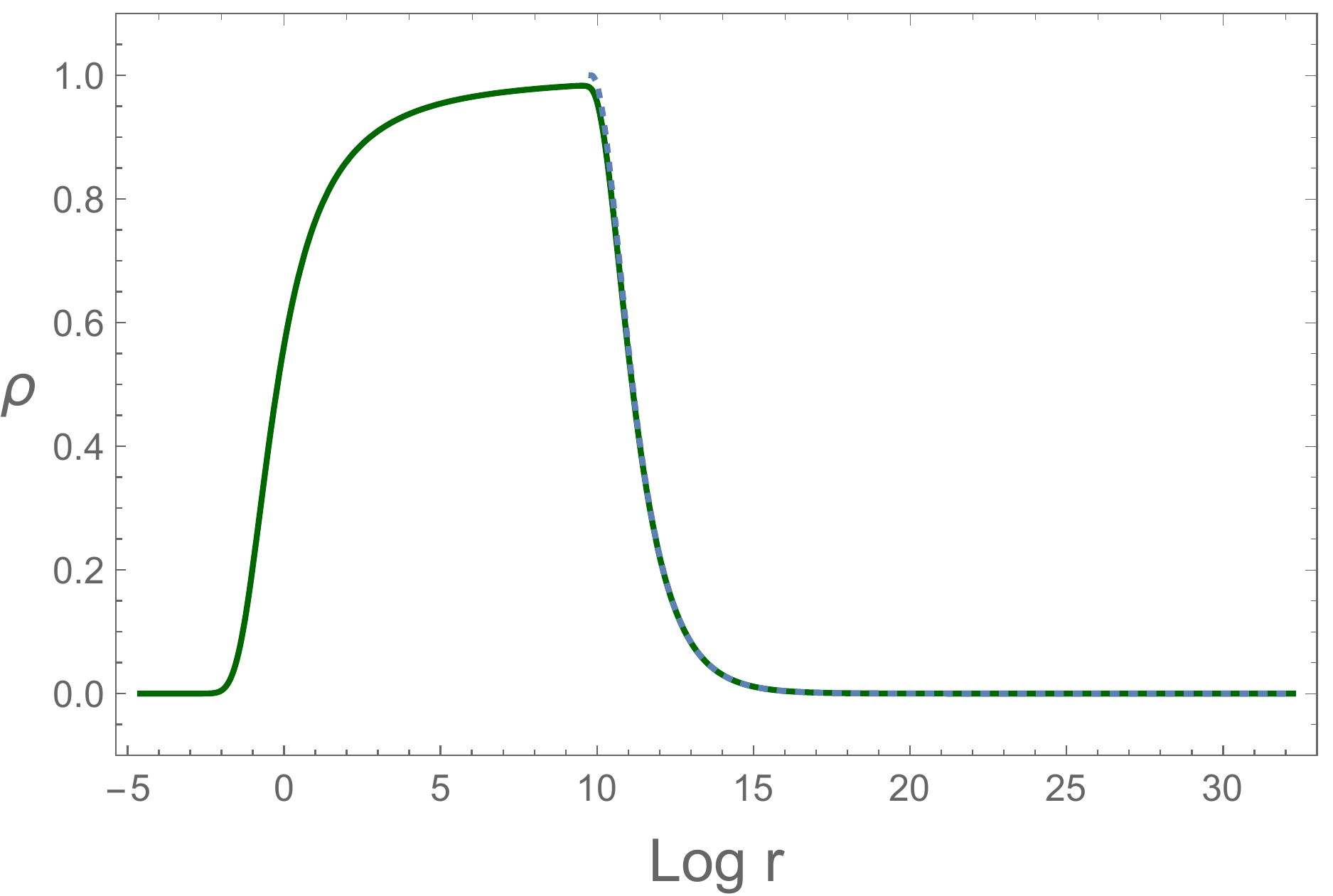}
\includegraphics[scale=0.35]{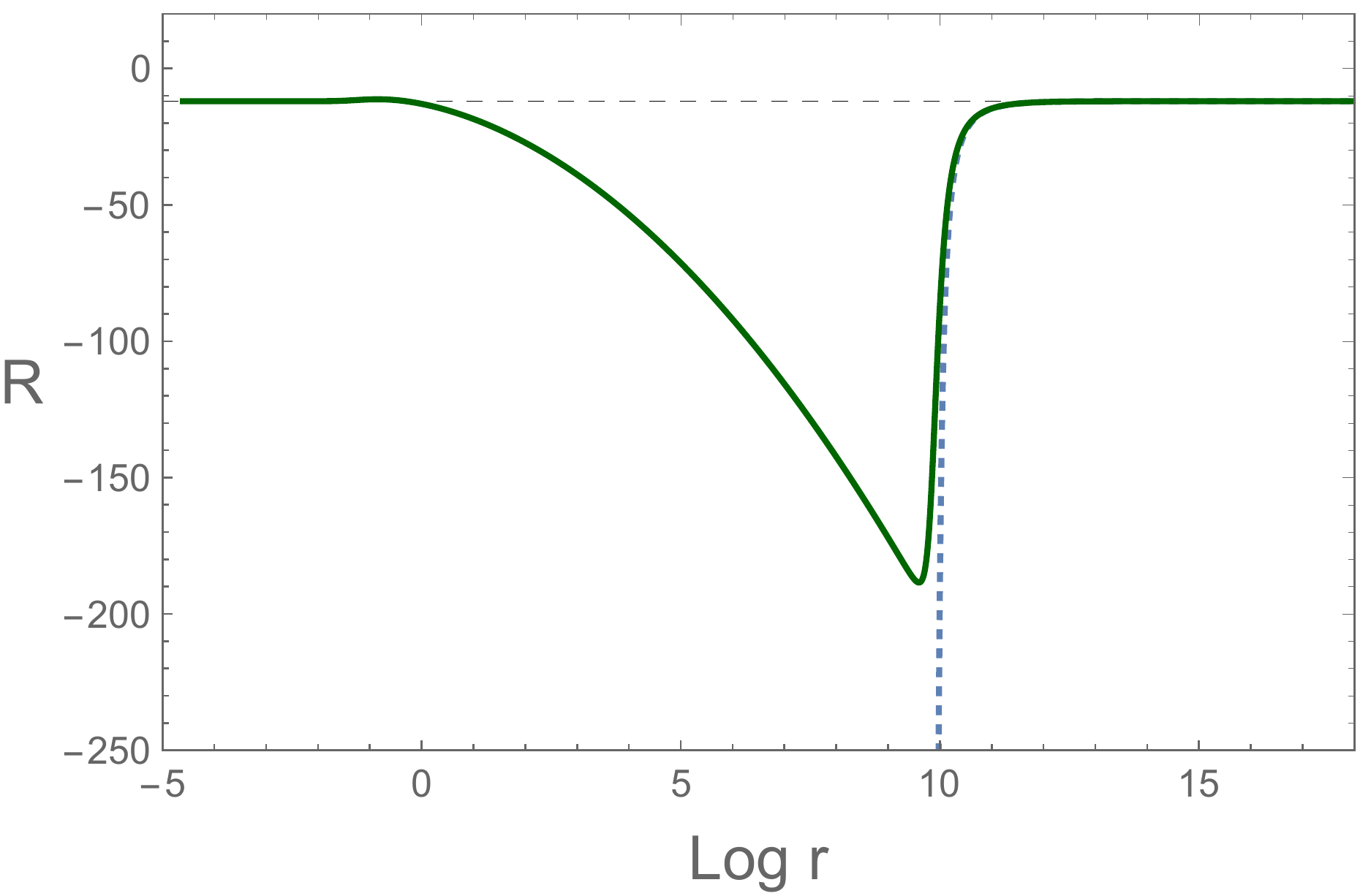}
\caption{\label{fig:bigAniso} The scalar profile, $\rho$, and Ricci curvature scalar, $R$, for a non-perturbative anisotropic boomerang flow with $\Lambda/k \approx 34.2$. The light blue dotted line shows the dielectric flow solution whose leading fall-off matches that of the full boomerang flow. The dashed black line shows the unit radius $AdS_4$.}
\end{figure}

In figure \ref{fig:bigAniso} we have plotted some features of the boomerang RG flow for a specific large deformation,
$\Lambda/k\approx 34.2$. The figure shows the scalar field profile as well as the Ricci scalar as functions of the proper radial coordinate, $\log r$. We see that such large deformations  drive the scalar towards the edge of field space, the boundary of the Poincar\'e disk at $\rho \to 1$, before it dives back towards zero, associated with
the unit radius $AdS_4$. 
In both plots we additionally depict a dotted light blue line, which corresponds to the Poincar\'e invariant
dielectric solution, with the value of $\mu$ in \eqref{poincflowexp} chosen to agree with the leading order fall-off of the numerical solution at the $AdS_4$ boundary. 
Specifically, we take $2\mu=\Lambda\sim 1.8\times 10^4$ for this flow. 
Clearly, for large deformations the boomerang flows are closely tracking
the Poincar\'e invariant solution, as we expect.

In section \ref{pinvflow} we discussed how the Poincar\'e invariant RG flow has the feature that
generic operators will have spectral functions with a hard gap at low frequencies, $\omega\le 2\mu$.  
The behaviour of
the boomerang RG flows for large deformations, that we just described, 
allows us to infer how this gap is closed. Since for large values of $\log r$ the geometries
are nearly identical we can deduce that for large $\omega$ the spectral function for the boomerang flow will
closely approximate those of the Poincar\'e invariant flow given in \eqref{retgf}. However, as we approach $r\to \mu$, associated with $\omega \to 2\mu$, the geometry of the boomerang flow is modified from the Poincar\'e invariant flow as it
heads back to the $AdS_4$ solution in the IR. This implies that the hard gap is replaced with
a small bump of spectral weight in the region $0\le  \omega <2\mu$, in order that it approaches the
power law behaviour, $\omega^3$, dictated by conformal invariance\footnote{As an aside, for certain operators we note that the power of $\omega$ appearing in the spectral functions in the UV and IR are not necessarily the same, as discussed in
section 4.2 of \cite{Donos:2017ljs}.}
 in the far IR.

\section{Supersymmetric boomerang solutions in $D=11$}\label{seceleven}

We can uplift the $D=4$ Susy Q-lattice solutions to $D=11$ either using the formulae
for solutions of the $\mathcal{N}=2$ STU model in \cite{Cvetic:2000tb,Azizi:2016noi} or the formulae for the 
$\mathcal{N}=4$ $SO(4)$ gauged supergravity model in \cite{Cvetic:1999au}.

To use the results of \cite{Cvetic:1999au} we should switch from the complex scalar field $z$, parametrising the Poincar\'e
disc, to scalar fields $\phi,\chi$, which parametrise the upper half plane, as
described in appendix \ref{conkk}. For the Susy Q-lattice solutions in the ansatz \eqref{Susyqansatz} we have
\begin{equation}
e^\phi = \cosh\lambda+\sinh\lambda \cos kx\,, \qquad \chi e^\phi = \sinh \lambda\sin kx\,, \qquad \mathrm{with}\qquad \rho=\tanh\frac{\lambda}{2}\,.
\end{equation}
We commented before that we can take $kx\to kx+\theta$, where $\theta$ is a constant, 
and the Q-lattice solutions are otherwise unchanged. When
$k\ne 0$, we can remove $\theta$ by a shift of the $x$ coordinate. However, when $k=0$, the value of $\theta$ parametrises different solutions in $D=11$, as emphasised in \cite{Pope:2003jp}.

The uplifted $D=11$ metric can be written
\begin{equation}
\dd s^2_{11} = \left(Z\tilde{Z}\right)^{1/3}\dd s^2_4+4\left(Z\tilde{Z}\right)^{1/3}\left[\dd\xi^2+\frac{\cos^2\xi}{Z}\dd\Omega_3^2+ \frac{\sin^2\xi}{\tilde{Z}}\dd\tilde{\Omega}_3^2\right]\,,
\end{equation}
where $\dd\Omega_3^2$ and $\dd\tilde{\Omega}_3^2$ are each metrics on a unit radius, round three-sphere,
and 
\begin{align}
Z &= \sin^2\xi + \cos^2\xi\left( \cosh\lambda+\cos kx\sinh\lambda\right)\,,\nn
 \tilde{Z} &= \cos^2\xi + \sin^2\xi\left( \cosh\lambda-\cos kx\sinh\lambda\right)\,.
\end{align}
The four-form flux can be written as the sum of three terms
\begin{equation}
F_{(4)} = F_{(4)}^{FR}+F_{(4)}^{I_1} + F_{(4)}^{I_2}.
\end{equation}
The `Freund-Rubin' part of the flux, $F_{(4)}^{FR}$, is given by
\begin{equation}
F_{(4)}^{FR} = -\big(2+\cosh\lambda+\cos 2\xi\sinh\lambda\cos kx \big)\mathrm{vol}_4 \,,
\end{equation}
where $\mathrm{vol}_4$ is the volume form for $ds^2_4$.
The remaining components of the flux are given by
\begin{equation}
F_{(4)}^{I_1} = \sin 2 \xi \Big(-\cos k x\star\dd \lambda+\frac{1}{2} \sin k x \sinh 2\lambda\star \dd (kx)\Big)\wedge \dd\xi\,,
\end{equation}
where $\star$ 
is the hodge dual is with respect to $\dd s^2_4$,
and 
\begin{align}
F_{(4)}^{I_2} = \dd\left(\sinh\lambda\sin kx\frac{\cos^4\xi}{Z}\right)\wedge \mathrm{vol}_{\Omega_3}
-
\dd\left(\sinh\lambda\sin kx\frac{\sin^4\xi}{\tilde Z}\right)\wedge \mathrm{vol}_{\tilde \Omega_3}\,.
\end{align}

In the dielectric RG flows of \cite{Pope:2003jp}, obtained by setting $kx\to\theta$ for constant $\theta$ in the
above, the $D=11$ solutions were interpreted as a distribution of membranes that have
polarized into fivebranes wrapping the three-spheres. For the boomerang flows this distribution is
spatially modulated in the $x$ direction. In particular, the number of membranes is fixed
by integrating $\star F_4$ over the seven-sphere at the $AdS_4$ boundary at $r\to\infty$ where 
$\lambda(r)\to 0$.

\section{Spatially modulated mass deformations in ABJM theory}\label{secabjm}
ABJM theory is a $d=3$ Chern-Simons theory coupled to matter with gauge group $U(N)_q \times U(N)_{-q}$, where $q$ labels the Chern-Simons level on each factor \cite{Aharony:2008ug}. 
It is an interacting SCFT that describes the low energy dynamics of $N$ M2 branes on $\mathbb{C}^4/\mathbb{Z}_q$. It has manifest 
$\mathcal{N}=6$ supersymmetry and $SU(4)\times U(1)_b$ global symmetry.  When $q=1,2$ there is an enhancement of supersymmetry to $\mathcal{N}=8$. In the large $N$ and strongly coupled limit, the physics of the ABJM theory is captured holographically by $D=11$ supergravity on $AdS_4\times S^7/\mathbb{Z}_q$
with $\mathbb{Z}_q\subset U(1)_b\subset SO(8)$.

The supersymmetric boomerang RG flows that we have constructed lie within the 
$SO(4)\times SO(4)\subset SO(8)$ invariant sector
of $\mathcal{N}=8$ gauged supergravity. In particular, the uplifted $D=11$ solutions in the last section survive 
a cyclic quotient of the $S^7$ by $\mathbb{Z}_q\subset U(1)_b\subset SO(8)$
and then describe RG flows associated with spatially modulated deformations
of $\mathcal{N}=6$ ABJM theory. We now want to identify which operators in
ABJM field theory correspond to these deformations. A related analysis, for a different problem, appears
in \cite{Freedman:2013ryh} and we refer to that reference for additional details.

The $\mathbf{35}_v$ scalars and $\mathbf{35}_c$ psuedoscalars of $\mathcal{N}=8$ gauged supergravity
are conveniently parametrised in the 56-vielbein $\mathcal{V}$ in `unitary gauge' via  
\begin{equation}
\mathcal{V} = 
\begin{pmatrix}
u_{ij}\,^{IJ} & v_{ijKL}\\
&\\
v^{klIJ}& u^{kl}\,_{KL}
\end{pmatrix}
= \exp
\begin{pmatrix}
0 & \Sigma \\
\Sigma^* & 0
\end{pmatrix}\,,
\end{equation}
where $\Sigma$ satisfies $\Sigma_{I_1\dots I_4}=\frac{1}{4!}\epsilon_{I_1\dots I_8}\Sigma^{I_5\dots I_8}$
with $\Sigma^{IJKL}=\Sigma_{IJKL}^*$, and the $I$ index is an $\mathbf{8}_s$ index.
If we take the
$SO(4)\times SO(4)$ action to be rotating the $1234$ and $5678$ indices of $\mathbf{8}_s$, 
we see that we can identify our supergravity scalar $z$ with $\Sigma_{1234}=\Sigma^*_{5678}$.

To identify the dual operator in ABJM theory, we recall that the field content
includes bosons, $Y^A$, and fermions, $\psi_A$ transforming under 
$SU(4)\times U(1)_b$ in the $\bf{4}_0$ and $\bar{\bf{4}}_0$, respectively. 
From these we can form the following operators, each transforming in the $\mathbf{15}_0$, schematically of
the form
\begin{equation}\label{eq:150}
\mathcal{O}^{\Delta = 1} \sim \mathrm{Tr}\left(Y^A Y^\dagger_B - \frac{1}{4}\delta^A_B\, Y\cdot Y^\dagger  \right), \qquad \mathcal{O}^{\Delta = 2} \sim \mathrm{Tr}\left(\psi^\dagger\,^A\psi_B - \frac{1}{4}\delta^A_B\, \psi^\dagger\cdot \psi  \right)\,,
\end{equation}
and with scaling dimensions $\Delta = 1,2$, respectively. We will not nail down precise normalisations
of the operators and we also note that the $\Delta=2$ operator is supplemented by additional
terms quartic in the $Y$'s, as dictated by the supersymmetry algebra \cite{Gomis:2008vc}.

Returning to supergravity, under the decomposition $SU(4)\times U(1)_b\subset SO(8)$ we have the
branchings
\begin{align}
\mathrm{Re}\,\Sigma_{IJKL}&:\qquad  \mathbf{35}_v\to 
\mathbf{15}_0 +\mathbf{10}_2 +\overline{\mathbf{10}}_{-2} \,,\nn
\mathrm{Im}\,\Sigma_{IJKL}&:\qquad \mathbf{35}_c \to\mathbf{15}_0 +\mathbf{10}_{-2} +\overline{\mathbf{10}}_{2} \,,
\end{align}
and we can identify the $\mathbf{15}_0$'s of $\mathrm{Re}\,\Sigma_{IJKL}$ and $\mathrm{Im}\,\Sigma_{IJKL}$
with the above operators with $\Delta = 1,2$, respectively. To identify the operator associated with the
$SO(4)\times SO(4)$ singlet, $\Sigma_{1234}$, we can consider the common embedding of the subgroup
$U(1)\times U(1)_b\times SU(2)\times SU(2)\sim U(1)\times U(1)_b\times SO(4)$ into both
$U(1)_b\times SU(4)$ and $SO(4)\times SO(4)$. We take the
$U(1)$ to rotate the $12$ indices of $\mathbf{8}_s$, $U(1)_b$ to rotate the $34$ indices and, as before, the $SU(2)\times SU(2)\sim SO(4)$ rotates the $5678$ indices. Clearly, $\Sigma_{1234}$ is a singlet under 
$U(1)\times U(1)_b\times SU(2)\times SU(2)$. The associated singlet operator can easily be identified if we
take the two $SU(2)$ factors to act on $A=1,2$ and $A=3,4$ respectively, with $U(1)_b$ acting as an overall phase and the two doublets having opposite charge under $U(1)$ \cite{Freedman:2013ryh}. In particular we conclude that we can make the following identification between the real and imaginary
parts of the supergravity fields and the ABJM operators:
\begin{align}
\mathcal{X} \quad&\, \longleftrightarrow \quad\mathcal{O}_\mathcal{X}^{\Delta=1} \sim M_A\,^B \mathrm{Tr} \left(Y^A Y^\dagger_B\right)\,,\nn
\mathcal{Y} \quad&  \longleftrightarrow \quad \mathcal{O}_\mathcal{Y}^{\Delta=2} \sim M_A\,^B\mathrm{Tr} \left(\psi^\dagger\,^A  \psi_B \right)\label{eq:CFTOP}\,,
\end{align}
where $M_A\,^B = \mathrm{diag}(1,1,-1,-1)$ and we recall that in the Susy Q solutions we have 
$\mathcal{X}=\rho \cos kx $ and $\mathcal{Y}=\rho \sin kx $.

In appendix \ref{appb} we show that as solutions of $\mathcal{N}=8$ gauged supergravity, the
Susy Q boomerang flows preserve 8 supersymmetries.  After taking the $\mathbb{Z}_q$
quotient associated with ABJM theory, 6 of these supersymmetries survive for $q>2$. 
Thus, the spatially modulated deformation breaks the $\mathcal{N}=6$ 
supersymmetry of ABJM theory to $\mathcal{N}=3$ for $q>2$.
In appendix \ref{appb} we show that the 6 preserved supersymmetries can
be characterised as four Majorana spinors with eigenvalue $+1$ under the action of the $d=3$ gamma matrix
$\gamma^{\hat x}$ and two Majorana spinors with eigenvalue $-1$. In the special case of $q=1,2$ the supersymmetry is broken from 
$\mathcal{N}=8$ to $\mathcal{N}=4$ characterised by four plus four Majorana spinors with the previous stated
eigenvalues.

In a recent work it was shown that mass terms for ABJM, depending on 
one of the spatial coordinates, can preserve $\mathcal{N}=3$ supersymmetry in \cite{Kim:2018qle}.
Our results provide a precise holographic realisation of a specific example of these deformations. In the notation of 
\cite{Kim:2018qle}, the spatially dependent mass terms in the Lagrangian can be written as a sum of three terms:
\begin{align}\label{kimetal}
\Delta\mathcal{L}=m'M_A{}^B\mathrm{Tr} \left(Y^A Y^\dagger_B\right)
+mM_A{}^B\mathrm{Tr} \left(i\psi^\dagger\,^A \psi_B+\frac{8\pi}{q}Y^CY^\dagger_{[C}Y^A Y^\dagger_{B]}\right)
+m^2\mathrm{Tr} \left(Y^A Y^\dagger_A\right)
\end{align}
where $m$ is an arbitrary function of one of the spatial coordinates. To compare with
our holographic construction we should take the function $m(x)\propto \sin kx$. Up to the normalisation
of the operators coming from holography, which we haven't made precise, we see that the first two terms
in \eqref{kimetal} agree with our results (after recalling that the term quartic in $Y$'s is needed for the supersymmetric completion of the $\Delta=2$ operator, as mentioned above). The final term in
\eqref{kimetal} is an unprotected operator and hence is not visible from the supergravity point of view.

In \cite{Kim:2018qle} it was also determined how the supersymmetry is broken from $\mathcal{N}=6$ to 
$\mathcal{N}=3$ and we find agreement with the projections given in eq. (2.20) of \cite{Kim:2018qle}.
An interesting point is that our supergravity analysis shows that for the special case that $q=1,2$, when
ABJM has enhanced $\mathcal{N}=8$ supersymmetry, the spatially modulated deformation will preserve
$\mathcal{N}=4$ supersymmetry, a point that cannot be seen from the field theory construction of
\cite{Kim:2018qle}. Given that we have shown we get the supersymmetry enhancement for the special case
that $m(x)\propto \sin kx$, it is natural to conjecture that it will hold for arbitrary $m(x)$.

\section{Conclusion}
We have shown that simple Susy Q constructions are possible within the context of $\mathcal{N}=1$
supergravity in $D=4$. The constructions we considered are spatially anisotropic; based on the analysis
of general supersymmetric solutions of \cite{Gran:2008vx}, 
it seems likely that this is a general restriction for Susy Q for
these supergravity theories.

For $\mathcal{N}=1$ supergravity coupled to
a single chiral field, our construction required that the superpotential was constant in addition
to the standard Q-lattice restriction that the action has a global symmetry.
Although restrictive, this class includes a particularly interesting top-down example, arising
from the $SO(4)\times SO(4)\subset SO(8)$ invariant sector of $\mathcal{N}=8$ gauged supergravity
and hence of relevance to ABJM theory. It would be interesting to know if there
are other consistent truncations with constant superpotentials. Within
$\mathcal{N}=8$ gauged supergravity there is a classification of truncations of $\mathcal{N}=8$ gauged supergravity
that keeps scalars parametrising $SL(2)/SO(2)$, that are invariant under a group $G\subset SO(8)$ 
\cite{Bobev:2013yra}. There are three examples. In addition to the case
with $G=SO(4)\times SO(4)$, studied in this paper, there are also cases with 
$G=SU(3)\times U(1)\times U(1)$ and $G =G_2$. While in these latter two cases
the truncated action has an $SO(2)$ global symmetry that can be utilised for Q-lattice constructions,
in neither case is the superpotential constant. Of course this does not rule out the possibility that
there are other consistent truncations of $D=10,11$ supergravity to $D=4$ that do allow Susy Q constructions.

The Susy Q solutions that we explicitly constructed for the $SO(4)\times SO(4)$ invariant sector of $\mathcal{N}=8$ gauged supergravity are boomerang RG flows. They start at the ABJM vacuum in the UV, deformed by spatially modulated operators, and then flow in the IR back to the ABJM vacuum. We identified the participating operators in
the ABJM field theory and showed that it was consistent with the recent work of \cite{Kim:2018qle} who
considered spatially dependent mass deformations of ABJM theory, parametrised by an arbitrary function $m(x)$, preserving $\mathcal{N}=3$
supersymmetry for Chern-Simons level $q>2$. An interesting corollary of our work is that for $q=1,2$
there is an enhancement of supersymmetry from $\mathcal{N}=3$ to $\mathcal{N}=4$. The mass deformations
captured by the Susy Q construction have spatial dependence of the form $m(x)\propto \sin kx$.
It would be interesting to construct the gravity solutions for arbitrary $m(x)$ within the $SO(4)\times SO(4)$ invariant sector of $\mathcal{N}=8$ gauged supergravity. 
The supersymmetric Janus solutions of \cite{DHoker:2009lky,Bobev:2013yra} provide another example of $m(x)$, with
$m(x)\propto \delta(x)$, which can be found by solving ODEs. In general, however, one will need to solve PDEs. 
More generally, it would be interesting
to determine the most general spatially dependent and supersymmetric deformations of ABJM theory, both from the perspective of the field
theory and $D=11$ supergravity.

We calculated the index of refraction in the direction in which translations are broken, $n_x$, for the
Susy Q boomerang flows and showed that $n_x>1$. 
Combined with analogous results for a class of isotropic Q-lattices \cite{Donos:2017sba} and also for
the perturbative inhomogeneous solutions found in \cite{Chesler:2013qla}, 
we expect that this is a general result in holography.
It would be interesting to know if it is also true for boomerang RG flows in field theory more generally.

It would also be interesting to investigate Susy Q constructions in the context of supergravity theories in other
spacetime dimensions.
In fact, a construction in the context of a specific top-down $D=5$ gauged supergravity was already made in \cite{Donos:2014eua} which involves two axion-fields, with shift symmetries, that each depend linearly on one of two different spatial directions. 
That translations in two of the three spatial directions of the CFT are broken is associated with the fact
that there is a projection on the Killing spinor involving the time and the remaining spatial direction, analogous
to \eqref{ksbeg}. A difference to the construction of this paper, however,
is that the axions are dual to marginal operators in the dual CFT, rather than relevant operators. Furthermore, and
interrelated with this point, the $D=5$ solutions of \cite{Donos:2014eua} are not boomerang RG flows, but instead 
flow
from $AdS_5$ in the UV to an $AdS_3\times\mathbb{R}^2$ fixed point in the IR, suported by the two linear axions. 
It seems likely that Susy Q solutions in $D=5$ using fields dual to relevant operators can also be found.

\section*{Acknowledgements}
We thank Aristomenis Donos for collaboration on related material and also Louise Anderson,
Igal Arav, Carlos N\'u\~nez, Matthew Roberts and Toby Wiseman for
helpful discussions.
JPG and CR are supported by the European Research Council under the European Union's Seventh Framework Programme (FP7/2007-2013), ERC Grant agreement ADG 339140. JPG is also supported by STFC grant ST/P000762/1, EPSRC grant EP/K034456/1, as a KIAS Scholar and as a Visiting Fellow at the Perimeter Institute.  

\appendix

\section{Useful results for $\mathcal{N} = 1$ supergravity}\label{sec:Nis1con}
Here we collect a few useful formulae, mostly using the conventions given in \cite{fvpbook}.

We begin by considering $\mathcal{N}=1$ supergravity in $D=4$ coupled to an arbitrary number of chiral multiplets with the bosonic part of the
action given by 
\begin{equation}\label{eq:Sgfirstone}
S_G = \int  \dd^4 x \sqrt{-g}\Big(R - G_{\alpha\bb}\partial_\mu z^\alpha \partial^\mu \bar{z}^{\bb}-\mathcal{V}\Big) \,.
\end{equation}
The complex scalar fields $z^\alpha$ parametrise a K\"ahler manifold with metric given by
\begin{equation}\label{eq:Nis1data}
G_{\alpha\bb} = 2\partial_\alpha\partial_{\bb}\mathcal{K}\,,
\end{equation}
where $\mathcal{K}$ is the K\"ahler potential\footnote{Note that we have set $16\pi G=2\kappa^2=1$ and
our K\"ahler potential, $\mathcal{K}$, is related to the one in \cite{fvpbook}, $\mathcal{K}_{there}$, via
$\mathcal{K}=\kappa^2 \mathcal{K}_{there}=(1/2)\mathcal{K}_{there}$.}
 . The potential $\mathcal{V}$ is given by
\begin{equation}\label{potterms}
\mathcal{V} = 4 G^{\alpha\bb}\partial_\alpha\mathcal{W}\partial_{\bb}\mathcal{W}-\frac{3}{2}\mathcal{W}^2\,,\qquad
\mathcal{W} =-e^{\mathcal{K}/2}|W|\,,
\end{equation}
where $W$ is the superpotential which is a holomorphic function of the $z^\alpha$.

We use gamma matrices given by
\begin{equation}
\Gamma^{\hat \mu} = 
\begin{pmatrix}
0 & \sigma^{\hat \mu} \\
\bar{\sigma}^{\hat \mu} & 0
\end{pmatrix}\,,
\qquad \mathrm{with}\qquad \sigma^{\hat \mu} = (1,\vec{\sigma}), \qquad \bar{\sigma}^{\hat \mu} = (-1, \vec{\sigma})\,,
\end{equation}
where the hatted indices refers to an orthonormal frame,
and $\vec{\sigma}$ are the three Pauli matrices. The supersymmetry parameter $\hat{\epsilon}$ satisfies a Majorana condition which can be expressed in terms of chiral components via
\begin{equation}
\hat{\epsilon} = (\epsilon,\tilde{\epsilon})^T
\qquad \mathrm{with} \qquad \tilde{\epsilon} = i\sigma_{2} \,\epsilon^*\,,
\end{equation}
and the chiral projections are given by
\begin{equation}
\Gamma^5 = -i \Gamma^{\hat 0}\Gamma^{\hat 1}\Gamma^{\hat 2}\Gamma^{\hat 3}, \qquad P_{L/R} = \frac{1}{2}\left(1\pm\Gamma^5 \right)\,.
\end{equation}

The variations for the fermions as chiral spinors can then be written as
\begin{align}\label{Susyvars}
\delta \psi_\mu & =\, \left(\nabla_\mu -\frac{3}{2}i\mathcal{A}_\mu \right) \epsilon + \frac{1}{4}\sigma_\mu e^{\mathcal{K}/2} W \tilde{\epsilon}\,,\nn
\delta\tilde{ \psi}_\mu & =\, \left(\tilde{\nabla}_\mu +\frac{3}{2}i\mathcal{A}_\mu \right) \tilde{\epsilon} + \frac{1}{4}\bar{\sigma}_\mu e^{\mathcal{K}/2} \bar{W} \epsilon\,,\nn
\sqrt{2}\delta\chi^\alpha & = \,\sigma^\mu\partial_\mu z^\alpha\tilde{\epsilon} -e^{\mathcal{K}/2}G^{\alpha\bb}D_{\bb}\bar{W}\epsilon \,,\nn
\sqrt{2}\delta\tilde{\chi}^{\bb} & = \,\bar{\sigma}^\mu\partial_\mu \bar{z}^{\bb}\epsilon - e^{\mathcal{K}/2}G^{\alpha\bb}D_{\alpha}W\tilde{\epsilon}\,,
\end{align}
where the various covariant derivatives are defined as
\begin{equation}
\nabla_\mu = \partial_\mu +\frac{1}{4}\omega_\mu\,^{\hat\nu\hat\rho}\sigma_{[\hat\nu}\bar{\sigma}_{\hat\rho]}, \qquad \tilde{\nabla}_\mu = \partial_\mu +\frac{1}{4}\omega_\mu\,^{\hat\nu\hat\rho}\bar{\sigma}_{[\hat\nu}\sigma_{\hat \rho]}, \qquad D_\alpha = \partial_\alpha +\partial_\alpha\mathcal{K},
\end{equation}
and the K\"ahler connection is given by
\begin{equation}
\mathcal{A}_\mu = \frac{i}{6}\sum_\alpha (\partial_\alpha\mathcal{K}\partial_\mu z^\alpha
-\partial_{\bar \alpha}\mathcal{K}\partial_\mu \bar z^{\bar \alpha})\,.
\end{equation}

\subsection{Supersymmetry for Susy Q}
We now restrict to the case of a single chiral multiplet with complex scalar, $z$ (dropping the index). 
We choose an orthonormal frame with $\vec{\sigma}=(\sigma^{\hat x},\sigma^{\hat y},\sigma^{\hat r})$.
We next substitute the
Q-lattice ansatz given in \eqref{Susyqansatz} into the supersymmetry variations \eqref{Susyvars} and impose
the projections
\begin{align}
\Gamma^{\hat r}\hat\epsilon=-\hat\epsilon,\qquad
\Gamma^{\hat t \hat y}\hat \epsilon=-\hat\epsilon\,,
\end{align}
or equivalently 
\begin{equation}\label{eq:proj}
\sigma^{\hat{r}}\tilde{\epsilon} = -\epsilon\,, \qquad \mathrm{and} \qquad  
\sigma^{\hat{x}}\tilde{\epsilon} = -i\epsilon\,,
\end{equation}
which also implies $\sigma^{\hat{y}}\epsilon = -\epsilon$.
This then leads to the BPS equations given in \eqref{bpsgen},
provided that the superpotential $W$ is taken to be constant (in order that we can divide out the $e^{ikx}$ factors) 
and real. If we wanted to work with a non-real $W$, by carrying out a K\"ahler transformation,
we can soak up the phase of $W$ with a phase
rotation of the Killing spinors, leading to the same BPS equations in \eqref{bpsgen} with $W$ replaced with $|W|$. We also obtain a radial equation for the Killing spinor which,
using the other BPS equations, can be solved as in \eqref{ksrad}.

We can also make a connection with the general analysis of supersymmetric solutions given in
\cite{Gran:2008vx}. Let $(e^{\hat t}, e^{\hat y},e^{\hat x},e^{\hat r})=(e^{A}dt, e^A dy, e^V dx, N dr)$ be an orthonormal frame
and define the complex one-form $e^{\hat 1}=-\frac{1}{\sqrt{2}}(e^{\hat x}+ie^{\hat r})$. We then find that the first 
BPS equation in \eqref{bpsgen}
(with real, constant $W$), can be written in the form
\begin{align}
i\sqrt{2}(dz)_{\hat 1}=e^{\mathcal{K}/2}G^{-1}D_{\bar z} \bar W\,,
\end{align}
which can be compared with eq. (3.8) of \cite{Gran:2008vx}
(note the latter uses supersymmetry transformations as in \cite{fvpbook} with $2\kappa^2=2$).

\subsection{Consistent truncations of $D=11$ supergravity}\label{conkk}
There is a consistent Kaluza-Klein reduction of $D=11$ supergravity on $S^7$ to the
maximal $\mathcal{N}=8$, $SO(8)$ gauged supergravity in $D=4$. There is a further consistent truncation
of the latter to an $\mathcal{N}=2$, $U(1)^4$, gauged supergravity coupled to three vector multiplets, called the STU model \cite{Cvetic:1999xp,Cvetic:2000tb}.
After setting the gauge fields to zero, the STU model can be recast in the language of $\mathcal{N}=1$ supergravity (appendix A of \cite{Cabo-Bizet:2017xdr} has a discussion of different presentations of this theory).
Specifically, the STU model with vanishing gauge fields has
\begin{equation}\label{eq:nis1}
\mathcal{K} = -\sum_\alpha \log\left[2(1-|z^\alpha|^2)\right]\qquad \mathrm{and} \qquad W = 4\sqrt{2}\big(1+z^1z^2z^3 \big).
\end{equation}
In \cite{Donos:2017ljs} isotropic Q-lattice solutions were constructed for this theory with one
of the three complex scalars set to zero. 

There is another consistent truncation of $\mathcal{N}=8$ $SO(8)$ gauged supergravity to 
the $\mathcal{N}=4$ $SO(4)$ gauged supergravity of \cite{Das:1977pu}.
After setting the gauge fields to zero, we can recast this in the language of $\mathcal{N}=1$ supergravity.
In fact the Lagrangian is obtained from the STU model with vanishing gauge fields and setting $z^2=z^3=0$.
This leads to
\begin{equation}\label{eq:nis2}
\mathcal{K} = - \log\left[8(1-|z|^2)\right]\qquad \mathrm{and} \qquad W = 4 \sqrt{2},
\end{equation}
and the bosonic Lagrangian given in \eqref{eq:Sg2bulk}:
\begin{equation}\label{eq:Sg2bulkaptext}
S= \int \dd^4 x \sqrt{-g}\Big(R - \frac{2}{(1-|z|^2)^2}\partial_\mu z \partial^\mu \bar{z}
+\frac{2(3-|z|^2)}{1-|z|^2}\Big) \,.
\end{equation}
The coordinate $z$ parametrises the Poincar\'e disc.  
If we redefine the modulus of $z$ via 
$z=\tanh\lambda/2 e^{i\sigma}$ then we obtain
\begin{equation}\label{eq:Sg3}
S = \int \dd^4 x \sqrt{-g}\Big(R - \frac{1}{2}(\partial \lambda)^2
- \frac{1}{2}\sinh^2\lambda(\partial \sigma)^2
+2(2+\cosh\lambda)
\Big) \,.
\end{equation}
It is also useful to employ another field redefinition, which maps the Poincar\'e disc to the upper half plane.
Specifically, we write $z=\frac{1+i\tau}{(1-i\tau)}$ with $\tau=\chi+ie^{-\phi}$ which then puts the action in the form
\begin{equation}\label{eq:Sg4}
S = \int \dd^4 x \sqrt{-g}\Big(R - \frac{1}{2}(\partial \phi)^2
- \frac{1}{2}e^{-2\phi}(\partial \chi)^2
+4+2\cosh\phi+\chi^2e^\phi
\Big) \,.
\end{equation}
In this parametrisation the global $U(1)$ symmetry that we use to build the Q-lattice solutions is not
immediately self-evident. On the other hand it is useful in order to uplift the solutions to $D=11$ using the
results of \cite{Cvetic:1999au}.

\subsection{Holographic renormalisation}\label{holren}
We now consider the bosonic part of the supergravity action, including
the boundary terms, given by
\begin{equation}\label{eq:Sg}
S= \int \dd^4 x \sqrt{-g}\Big(R - G_{\alpha\bb}\partial_\mu z^\alpha \partial^\mu \bar{z}^{\bb}-\mathcal{V}\Big) 
+2\int  \dd^3 x\sqrt{-\gamma} K+S_{ct}+S_L\,.
\end{equation}
We have included the standard Gibbons-Hawking term, as well as a counter term,
$S_{ct}$, and $S_L$ is an additional boundary term which ensures that
the operators associated with the real and imaginary parts of the scalar fields 
have different scaling dimensions in the dual field theory, as required by supersymmetry
We follow the conventions and analysis of \cite{Cabo-Bizet:2017xdr} but we work in units 
with $16\pi G = 1$, and also the $AdS_4$ vacuum solution is assumed to have unit radius.

We use a radial Hamiltonian formalism and it will be sufficient for our purposes to consider
metrics of the form
\begin{equation}\label{eq:metric}
\dd s^2 = N(r)^2 \dd r^2  + \gamma_{ij}(r) \dd x^i \dd x^j\,,
\end{equation}
which greatly facilitates holographic renormalisation. The asymptotic boundary is normal to the (outward pointing) unit vector $n^\mu = (1/N, 0,0,0)$, and carries extrinsic curvature $K_{ij} =\frac{1}{2N}\partial_r\gamma_{ij}$.
Writing $z^\alpha=\mathcal{X}^\alpha+i\mathcal{Y}^\alpha$ the contribution to the momenta conjugate
to $\gamma_{ij}$, $\mathcal{X}^\alpha$ and $\mathcal{Y}^\alpha$ arising from 
the first two terms in (\ref{eq:Sg}) is 
\begin{align}
\pi^{ij} = &\, \sqrt{-\gamma}\Big(K \gamma^{ij}-K^{ij} \Big)\,,\nn
\pi^{\mathcal{X}}_\alpha = &\, -\frac{1}{N}\sqrt{-\gamma}\Big(G_{\alpha\bb}\partial_r\bar{z}^{\bb}+G_{\beta\bar{\alpha}}\partial_rz^{\beta}\Big)\,,\nn
\pi^{\mathcal{Y}}_\alpha = &\, -\frac{i}{N}\sqrt{-\gamma}\Big(G_{\alpha\bb}\partial_r\bar{z}^{\bb}-G_{\beta\bar{\alpha}}\partial_rz^{\beta}\Big)\,,
\end{align}
respectively,

These momenta need to be supplemented with contributions coming from the boundary counterterms used to renormalise the holographic theory. For the supergravity theories and their solutions of interest, the counterterm action can be written
\begin{equation}\label{eq:Sct}
S_{\mathrm{ct}} = 2\int \dd^3 x\sqrt{-\gamma}\,\mathcal{W}\,,
\end{equation}
where the function $\mathcal{W}$ was defined in \eqref{potterms}
and satisfies 
$\mathcal{V} = 4 G^{\alpha\bb}\partial_\alpha\mathcal{W}\partial_{\bb}\mathcal{W}-\frac{3}{2}\mathcal{W}^2$.
The addition of (\ref{eq:Sct}) to the first two terms in (\ref{eq:Sg}) renders the on-shell action finite, and defines a well posed Dirichlet problem for the bulk fields. It also shifts the canonical momenta: $\pi_\Phi\to \Pi_\Phi \equiv \pi_\Phi + \delta S_{\mathrm{ct}}/\delta \Phi$ where $\Phi \in \{\gamma_{ij},\mathcal{X}^\alpha,\mathcal{Y}^\alpha \}$.

The Dirichlet problem may not be the only consistent variational problem for a given bulk action. Depending on the details, Neumann or mixed boundary conditions may also be permissible. 
For the present work, we are interested in models which supersymmetry dictates that the scalars
$\mathcal{X}^\alpha$ are dual to operators with dimension $\Delta_\mathcal{X}=1$ and 
the pseudoscalars $\mathcal{Y}^\alpha$ are dual to operators with dimension $\Delta_\mathcal{Y}=2$
This can be achieved by performing a Legendre transformation of the renormalised bulk on-shell action with respect to the field $\mathcal{X}^\alpha$. This manoeuvre ensures that the transformed on-shell action is holographically dual to the field theory generating function, a functional of the sources for the dual operators. 

This can be achieved by taking $S_L$ to be
\begin{equation}\label{eq:SL}
S_L = \int \dd^3 x \sqrt{-\gamma}J_\alpha \mathcal{X}^\alpha \qquad \mathrm{with}\qquad J_\alpha \equiv -\frac{1}{\sqrt{-\gamma}}\Pi_\alpha^\mathcal{X}\,.
\end{equation}
With this action the field theory sources for the scalar operators in the dual field theory are then
\begin{equation}\label{eq:defsfirst}
{\mathcal{Y}}_s^\alpha = \lim_{r\to\infty} r\mathcal{Y}^\alpha\,, \qquad 
\mathcal{X}_s^\alpha = \lim_{r\to\infty} r^2 J_\alpha\,,
\end{equation}
with boundary metric (source) given by $\lim_{r\to\infty}\frac{1}{r^2}\gamma_{ij}$. The associated 
one point functions are given by
\begin{align}
\langle\mathcal{T}^{ij} \rangle = &\,  \lim_{r\to\infty} r^5 \left(\frac{2}{\sqrt{-\gamma}}\Pi^{ij} +J_\alpha\mathcal{X}^\alpha \gamma^{ij}\right)\,,\nn
\langle \mathcal{O}_\mathcal{Y}^\alpha\rangle = &\,  \lim_{r\to\infty}r^2\left( \frac{1}{\sqrt{-\gamma}} \Pi^\mathcal{Y}_\alpha\right)\,,\nn
\langle  \mathcal{O}^\alpha_\mathcal{X}\rangle = &\,  \lim_{r\to\infty} r\mathcal{X}^\alpha\,,
\label{eq:vevs}
\end{align}
where we have chosen the lapse function $N = 1/r$ and additionally assume that the scalars vanish near the boundary as
\begin{equation}
z^\alpha \approx \frac{A^\alpha}{r}+\frac{B^\alpha}{r^2 }+ \ldots\,.
\end{equation}
The associated Ward identities are given by
\begin{align}\label{eq:Ward1}
 \partial^i\langle \mathcal{T}_{ij}\rangle  = &\, \langle  \mathcal{O}_{\mathcal{X}}\rangle\cdot\partial_j\mathcal{X}_s+ \langle  \mathcal{O}_{\mathcal{Y}}\rangle\cdot \partial_j\mathcal{Y}_s\,,\nn
 \langle \mathcal{T}^i\,_i\rangle = & \, (3-\Delta_\mathcal{X})\, \langle \mathcal{O}_{\mathcal{X}}\rangle\cdot\mathcal{X}_s+(3-\Delta_\mathcal{Y})\, \langle  \mathcal{O}_{\mathcal{Y}}\rangle\cdot \mathcal{Y}_s\,,
\end{align}
where we have used $\cdot$ to denote a sum over the scalar index $\alpha$.

Focussing on the bulk action considered in section \ref{sfthy}:
\begin{equation}\label{eq:Sg2bulkap}
S= \int\dd^4 x \sqrt{-g}\Big(R - \frac{2}{(1-|z|^2)^2}\partial_\mu z \partial^\mu \bar{z}
+\frac{2(3-|z|^2)}{1-|z|^2}\Big)\,,
\end{equation}
we have
\begin{align}
S_{ct}&=-4\int \dd^3 x \sqrt{-\gamma}(1+\frac{1}{2}|z|^2)\,,
\end{align}
where we have just kept the terms quadratic in the scalars (as higher powers do not contribute),
and
\begin{align}
S_L&=4\int \dd^3 x \sqrt{-\gamma}\left(r\mathcal{X}\partial_r\mathcal{X}
+
{\mathcal{X}^2}\right)\,.
\end{align}
We note that $S_L$ breaks the bulk global symmetry which rotates $z$ by a constant phase, and gives 
rise, in particular, to a stress tensor for Q-lattice solutions which is spatially modulated as in \eqref{vevs}.

\section{Supersymmetry in $\mathcal{N}=8$ gauged supergravity and ABJM}\label{appb}
The Susy Q solutions can be easily embedded in $\mathcal{N}=8$ $SO(8)$ gauged supergravity 
\cite{deWit:1981sst} by considering the sector of the latter that is invariant under $SO(4)\times SO(4) \subset SO(8)$ \cite{Cvetic:1999au}. 
Here we sketch out a few details of how to calculate the supersymmetry preserved in
the $\mathcal{N}=8$ theory.

We consider the following $SO(4) \times SO(4)$ invariant tensors
\begin{equation}
\Omega^L = \dd x^1\wedge\dd x^2\wedge\dd x^3\wedge\dd x^4\,, \qquad \Omega^R = \dd x^5\wedge\dd x^6\wedge\dd x^7\wedge\dd x^8\,,
\end{equation}
with the $x^I$ coordinates on $\mathbb{R}^8$. 
The bosonic sector of the relevant truncation of $\mathcal{N}=8$ follows from the scalar ansatz
\begin{equation}
\Sigma = \zeta\, \Omega^L + \bar{\zeta}\,\Omega^R\,,
\end{equation}
where the complex scalar $\zeta$ is related to the complex scalar $z$ of the $\mathcal{N}=1$ theory \eqref{eq:Sg2bulk}
by writing
$z = \rho e^{i\sigma}$ with $\rho=\tanh\frac{\lambda}{2}$ (see \eqref{eq:Sg3}) and defining
\begin{equation}
\zeta = \frac{\lambda}{4} e^{i\sigma}\,.
\end{equation}
The scalar/pseudoscalar fields of $\mathcal{N}=8$ gauged supergravity parametrise the non-compact
coset $E_{7(7)}/SU(8)$. The relevant coset representative for the truncation can be efficiently obtained from
$\Sigma$ by working in `unitary gauge', in which the 56-bein $\mathcal{V}$ takes the form
\begin{equation}
\mathcal{V} = 
\begin{pmatrix}
u_{ij}\,^{IJ} & v_{ijKL}\\
&\\
v^{klIJ}& u^{kl}\,_{KL}
\end{pmatrix}
= \exp
\begin{pmatrix}
0 & \Sigma \\
\Sigma^* & 0
\end{pmatrix}.
\end{equation}
To carry out the matrix exponentiation, it is helpful to define the projector
\begin{equation}
\Pi = \frac{1}{4}\left(\Omega^L\cdot \Omega^L + \Omega^R\cdot \Omega^R \right)\,, \qquad \mathrm{where} \qquad (A\cdot B)_{IJKL} \equiv \sum_{M,N} A_{IJMN} B_{MNKL}\,,
\end{equation}
which has the following nice properties:
\begin{equation}
\Pi\cdot\Pi = \Pi, \qquad  \Sigma^*\cdot\Sigma=\Sigma\cdot\Sigma^* = \frac{\lambda^2}{4}\, \Pi,\qquad 
\Sigma\cdot \Pi = \Sigma\,.
\end{equation}
Using these one obtains
\begin{equation}
u_{ij}\,^{IJ} = \delta_{ij}^{IJ} + \big(\cosh\frac{\lambda}{2} -1 \big)\Pi_{ijIJ}, \qquad v^{klIJ} = \frac{1}{2}\sinh\frac{\lambda}{2}\left(e^{-i\sigma}\Omega^L+e^{i\sigma}\Omega^R\right)_{klIJ}.
\end{equation}

Next, using standard formulae, which can be found in \cite{deWit:1981sst, Freedman:2016yue}, one can use these expressions to evaluate the various tensors that appear in the $\mathcal{N}=8$ supergravity Lagrangian and its supersymmetric variations. After some calculation, one eventually finds that the supersymmetry variations of the fermions can be written for eight left/right chiral spinor parameters $\epsilon^I$/$\epsilon_I$ as follows.
Breaking up the $SO(8)$ indices $I \in \{1,8\}$ into two sets of $SO(4)$ indices, such that 
$a,b,c,d\in \{1,\dots,4\}$ and $s,t,u,v\in\{5,\dots, 8\}$, we find
\begin{align}
\frac{1}{2}\delta\psi_\mu^a & = \,\left[ \nabla_\mu-\frac{1}{4}\left(\frac{\bar{z}\partial_\mu z - z \partial_\mu\bar{z}}{1-\bar{z}z}\right)\right]\epsilon^a+\frac{1}{2\sqrt{1-\bar{z}z}}\Gamma_\mu \epsilon_a\,,\nn
\frac{1}{2}\delta\psi_\mu^s & =\, \left[ \nabla_\mu+\frac{1}{4}\left(\frac{\bar{z}\partial_\mu z - z \partial_\mu\bar{z}}{1-\bar{z}z}\right)\right]\epsilon^s+\frac{1}{2\sqrt{1-\bar{z}z}}\Gamma_\mu \epsilon_s\,,
\end{align}
and
\begin{align}
\frac{1}{\sqrt{2}}\delta\chi^{abc} & = \,\frac{1}{1-\bar{z}z}\Gamma^\mu\partial_\mu\bar{z}\,\Omega^L\,_{abc}\,^d\epsilon_d-\frac{\bar{z}}{\sqrt{1-\bar{z}z}}\Omega^L\,_{abcd}\,\epsilon^d\,,\nn
\frac{1}{\sqrt{2}}\delta\chi^{stu} & = \,\frac{1}{1-\bar{z}z}\Gamma^\mu\partial_\mu z\,\Omega^R\,_{stu}\,^v\epsilon_v-\frac{z}{\sqrt{1-\bar{z}z}}\Omega^R\,_{stuv}\,\epsilon^v\,,
\end{align}
together with their conjugate variations. Evaluating these variations using the Susy Q ansatz \eqref{Susyqansatz}, one finds that the projections
\begin{equation}
\Gamma^{\hat{r}}\epsilon_a = -\epsilon^a\qquad \Gamma^{\hat{x}}\epsilon_a = i \epsilon^a\qquad\mathrm{and}\qquad\Gamma^{\hat{r}}\epsilon_s = -\epsilon^s, \qquad \Gamma^{\hat{x}}\epsilon_s = -i \epsilon^s\,,
\end{equation}
yield the same BPS equations as in (\ref{eq:bps1}). These projections reduce the supersymmetry from 32 to eight real components. If we form the Majorana spinors 
$\epsilon^I_{(M)} = \epsilon^I + \epsilon_I$ then we can write the projections as
\begin{equation}\label{bulkneight}
\Gamma^{\hat{r}}\epsilon^I_{(M)} = -\epsilon^I_{(M)}\,,\qquad \Gamma^{\hat{t}\hat{y}}\epsilon^a_{(M)}= +\epsilon^a_{(M)}\,,\qquad\Gamma^{\hat{t}\hat{y}}\epsilon^s_{(M)}= -\epsilon^s_{(M)}\,.
\end{equation}

\subsection{Supersymmetry in ABJM theory}

To understand the preserved supersymmetries in ABJM theory, it is convenient 
to use a different basis of Gamma matrices in $D=4$ than used in appendix \ref{sec:Nis1con}. 
Specifically, we can take
\begin{equation}
\Gamma^{\hat{\mu}} = (\Gamma^{\hat{i}},\Gamma^{\hat{r}}) \qquad \mathrm{where} \qquad \Gamma^{\hat{i}} = \gamma^{\hat{i}}\otimes \sigma^2, \qquad \Gamma^{\hat{r}} = -1\otimes \sigma^3\,,
\end{equation}
with, for example, $\gamma^{\hat{i}} = (i\sigma^3, -\sigma^2, -\sigma^1)$, so that $\gamma^{\hat{t}}\gamma^{\hat{x}}\gamma^{\hat{y}}=+1$.
In this basis, the bulk Killing spinors are of the form
\begin{equation}
\eta = \chi \otimes 
\begin{pmatrix}
1\\
0
\end{pmatrix},
\end{equation}
and hence $\gamma^{\hat{i}}$ are the $d=3$ Gamma matrices acting on 
the two component boundary spinor $\chi$. Furthermore, in this basis the bulk Majorana condition 
$\eta_{(M)} = B^{-1}\eta_{(M)}^*$ with $B = \sigma^1 \otimes \sigma^3$ becomes a constraint on the boundary spinor of the form $\chi_{(M)} = \sigma^1\chi_{(M)}^*$. In this language the projections on the bulk Killing
spinors given in \eqref{bulkneight} can be written in terms of the boundary spinors as
\begin{equation}\label{bulkneight2}
\gamma^{\hat{t}\hat{y}}\chi^a_{(M)}= +\chi^a_{(M)}\,,\qquad\gamma^{\hat{t}\hat{y}}\chi^s_{(M)}= -\chi^s_{(M)}\,,
\end{equation}
or equivalently
\begin{equation}\label{bulkneight3}
\gamma^{\hat{x}}\chi^a_{(M)}= -\chi^a_{(M)}\,,\qquad\gamma^{\hat{x}}\chi^s_{(M)}=+\chi^s_{(M)}\,,
\end{equation}
with $a\in \{1,\dots,4\}$ and $s\in\{5,\dots, 8\}$.

To describe the supersymmetries relevant to ABJM theory we need to examine which of the supersymmetries survive the cyclic quotient by $\mathbb{Z}_q\subset U(1)_b$ with $U(1)_b\times SU(4)\subset SO(8)$. 
Recall that in section \ref{secabjm} we took $U(1)_b$ to act on $\mathbf{8}_s$ as a rotation in the 34
components. We also recall that the eight gravitini of $\mathcal{N}=8$ gauged supergravity transform
as $\mathbf{8}_s$ and we have the branching $\mathbf{8}_s\to  \mathbf{6}_0 + \mathbf{1}_2 + \mathbf{1}_{-2}$.\
Thus for the generic case, the $\mathcal{N} = 6$ supersymmetries of ABJM theory at level $q>2$
are identified with $\chi^a_{(M)}$, with $a\in \{1,2\}$ and $\chi^s_{(M)}$ with $s\in\{5,\dots, 8\}$. The spatially modulated deformation breaks 1/2 of this supersymmetry to $\mathcal{N} = 3$, given by the projections above.
The projections as given in \eqref{bulkneight3} exactly correspond to those given in eq. (2.20) of \cite{Kim:2018qle}. For the special case of $q=1,2$ the non-manifest $\mathcal{N}=8$ supersymmetry of ABJM
will be broken to $\mathcal{N} = 4$, associated with the projections \eqref{bulkneight3} but now
with $a\in \{1,\dots,4\}$ and $s\in\{5,\dots, 8\}$.

\section{Isotropic boomerang RG flow}\label{sec:HRGs}

In \cite{Donos:2017ljs} isotropic boomerang RG flows were constructed in $D=11$ supergravity using
the consistent truncation to the $\mathcal{N}=2$ STU model in $D=4$ that we described in appendix \ref{conkk}.
Here we present the calculation of the stress tensor using the holographic renormalisation described in
appendix \ref{holren}.

Using a different radial coordinate to \cite{Donos:2017ljs}, the isotropic solutions lie within the ansatz
\begin{align}
ds^2&=-e^{2A}dt^2+e^{2V}(dx^2 + dy^2) +\frac{1}{r^2}dr^2\,,\nn
z^1 &= \rho e^{i k x}\,,\qquad z^2 = \rho e^{i k y}\,, \qquad z^3 = 0\,,
\end{align}
with $A$, $V$ and $\rho$ functions of $r$ only. The asymptotic expansions as $r\to\infty$ are
given by
\begin{align}
e^{2A} & = r^2 -{\dfone}^2 + M\frac{1}{r} + \ldots\,,\nn
e^{2V} & =\, r^2 -{\dfone}^2-(\frac{M}{2} +\frac{8}{3} {\dfone} {\dftwo})\frac{1}{r} + \ldots\,,\nn
\rho & =\, {\dfone}\frac{1}{r} + {\dftwo}\frac{1}{r^2}+\ldots\,.
\end{align}

Hence we deduce that the non-trivial field theory sources, with $\alpha=1,2$, are given by
\begin{equation}\label{eq:defs}
 \mathcal{X}^\alpha_s = -4{\dftwo}(\cos kx,\cos ky)\,,
 \qquad
{\mathcal{Y}}^\alpha_s = {\dfone}(\sin kx,\sin ky) \,,
\end{equation}
and the vevs are:
\begin{align}
\langle\mathcal{T}^{tt} \rangle = &\,-3 M+2 {\dfone} {\dftwo} (\cos 2 k x+\cos 2 k y-2) \,,\nn
\langle\mathcal{T}^{xx} \rangle = &\, \langle\mathcal{T}^{yy} \rangle= -\frac{3}{2} M
-2 {\dfone} {\dftwo} (\cos 2 k x+\cos 2 k y+2) \,,\nn
\langle \mathcal{O}_{\mathcal{X}^\alpha}\rangle = &\, {\dfone}(\cos k x,\cos ky)\,,\nn
\langle  \mathcal{O}_{\mathcal{Y}^\alpha}\rangle = &\,  4{\dftwo}(\sin kx, \sin ky)\,.
\end{align}
One can check that these correlation functions obey the expected Ward identities in (\ref{eq:Ward1}).

We observe that, in general, the spatial average of $\langle\overline{{\mathcal{T}}^{tt}} \rangle$ is non-zero.
Curiously, however, constructing a perturbative solution, as in \cite{Donos:2017ljs},
we find that at leading non-trivial order in the perturbative expansion it does vanish.

\section{Novel $AdS_{D-n}\times\mathbb{R}^n$ solutions}\label{appdee}
Consider a general class of theories in $D$ spacetime dimensions of the form
\begin{equation}\label{eq:actfirst}
S =\int\dd^Dx\,\sqrt{-g}\Big(R-V(\phi)-\frac{1}{2}G_{ab}(\phi)\partial\phi^a\partial\phi^b
-\frac{1}{2}G_{ij}(\phi)\partial\chi^i\partial\chi^j\Big)\,.
\end{equation}
Taking the index $i=1,\dots,n$ this theory has $n$ shift symmetries $\chi^i\to \chi^i+\epsilon^i$ which can
be used for Q-lattice constructions. We seek solutions of the form
\begin{align}
ds^2&=L^2ds^2(AdS_{D-n}) +h^2 dx^i dx^i\,,\nn
\phi^a&=\phi^a_0,\qquad \chi^i=k x^i\,,
\end{align}
where $ds^2(AdS_{D-n})$ has unit radius and $L^2$, $h$, $k$ and $\phi^a_0$ are all real constants.
We find that the equations of motion are satisfied if
\begin{align}
L^2=\frac{(D-2)(D-n-1)}{-V}\,,\qquad
{-V}\delta_{ij}=\frac{(D-2)k^2}{2h^2}G_{ij},\qquad
-\partial_aV=\frac{k^2}{2h^2}\delta^{ij}\partial_aG_{ij}\,,
\end{align}
with all quantities evaluated at the fixed point values of $\phi^a=\phi^a_0$.
Notice that $G_{ab}$ does not enter these conditions and also that $V(\phi^a_0)\ne 0$.

Let us now restrict to the sub-class of theories with $G_{ij}(\phi)=f(\phi)\delta_{ij}$, so that the equations we
need to solve are 
\begin{align}\label{redeqs}
L^2=\frac{(D-2)(D-n-1)}{-V}\,,\qquad
\frac{k^2}{h^2}=\frac{2}{D-2}\frac{-V}{f}\qquad
-\partial_aV=\frac{k^2n }{2h^2}\partial_af\,,
\end{align}
again evaluated at $\phi^a=\phi^a_0$. We have
also used the fact that since solutions require $V(\phi^a_0)\ne 0$, we must also have 
$f(\phi^a_0)\ne 0$.
We now notice the remarkable fact that if we consider theories in which we have, functionally,
\begin{align}
-V=c f^{\frac{n}{D-2}}\,,
\end{align}
where $c$ is a constant,
then the equations of motion are satisfied {\it for any value} of $\phi^a_0$ 
provided that $k^2/h^2$ and $L^2$ are given by the first two conditions in \eqref{redeqs}.
For example, if we restrict to a single scalar field, $\phi$, with $ f=e^{2\phi}$,
\begin{equation}\label{eq:act}
S =\int\dd^Dx\,\sqrt{-g}\Big(R-V(\phi)-\frac{1}{2}G(\phi)(\partial\phi)^2-\frac{1}{2}e^{2\phi}\sum_{i=1}^n(\partial\chi^i)^2\Big)\,,
\end{equation}
then for $D=4$ we have $AdS_{3}\times\mathbb{R}$ solutions for $V=-c e^{\phi}$ and
$AdS_{2}\times\mathbb{R}^2$ solutions for $V=-c e^{2\phi}$. For $D=5$ we have 
$AdS_{4}\times\mathbb{R}$ solutions for $V=-c e^{2\phi/3}$,
$AdS_{3}\times\mathbb{R}^2$ solutions for $V=-c e^{4\phi/3}$
and $AdS_{2}\times\mathbb{R}^3$ solutions for $V=-c e^{2\phi}$. 
Of course, for these theories there is no $AdS_4$ or $AdS_5$ vacuum solution.

If we consider the top-down $D=4$ model \eqref{eq:Sg2bulk}, or more conveniently in the form
\eqref{eq:Sg3}, then we see that as $|z|\to 1$, or equivalently, $\lambda\to \infty$,
the model is approximately of the above form with $V=-e^\lambda$ (after taking $\sigma\to2\chi$).
This approximate model thus has $AdS_3\times\mathbb{R}$ solutions with $\lambda_0$ unspecified. 
For large deformations, in the Susy Q boomerang RG flows the scalar field is becoming large and
it is natural to wonder if the bulk solutions have an intermediate regime approximated
by these $AdS_3\times\mathbb{R}$ solutions of the auxiliary theory. However, we do not see any direct
evidence for this in figure \ref{fig:bigAniso}. It seems likely that this is connected to
the fact that the scalar field is not fixed in the $AdS_3\times\mathbb{R}$ solutions.

\providecommand{\href}[2]{#2}\begingroup\raggedright\endgroup

\end{document}